\documentclass[a4paper,twocolumn,english,aps,prl,superscriptaddress]{revtex4-1}
\usepackage[T1]{fontenc}
\usepackage[latin9]{inputenc}
\usepackage{color}
\usepackage{babel}
\usepackage{amsmath}
\usepackage{amssymb}
\usepackage{graphicx}
\usepackage{esint}
\usepackage[unicode=true,pdfusetitle,
 bookmarks=true,bookmarksnumbered=false,bookmarksopen=false,
 breaklinks=false,pdfborder={0 0 1},backref=false,colorlinks=false]
 {hyperref}
\usepackage{breakurl}

\makeatletter

\usepackage{babel}
\usepackage{babel}

\makeatother

\begin{document}

\title{Dirac States in an Inclined Two-Dimensional Su-Schrieffer-Heeger Model}

\author{Chang-An Li}
\email{changan.li@uni-wuerzburg.de}

\affiliation{Institute for Theoretical Physics and Astrophysics, University of
Würzburg, 97074 Würzburg, Germany}

\author{Sang-Jun Choi}

\affiliation{Institute for Theoretical Physics and Astrophysics, University of
Würzburg, 97074 Würzburg, Germany}

\author{Song-Bo Zhang}

\affiliation{Department of Physics, University of Zürich, Winterthurerstrasse
190 8057, Zürich, Switzerland}

\author{Björn Trauzettel}

\affiliation{Institute for Theoretical Physics and Astrophysics, University of
Würzburg, 97074 Würzburg, Germany}

\affiliation{Würzburg-Dresden Cluster of Excellence ct.qmat, Germany}

\date{\today}
\begin{abstract}
We propose to realize Dirac states in an inclined two-dimensional Su-Schrieffer-Heeger
model on a square lattice. We show that a pair of Dirac points protected
by space-time inversion symmetry appear in the semimetal phase. 
The locations of these Dirac points are not pinned to any
high-symmetry points of the Brillouin zone but are tunable through
parameter modulations. Interestingly, the merging of two Dirac points undergoes
a topological phase transition that leads to either a weak
topological insulator or a nodal-line semimetal. We provide a systematic
analysis of these topological phases from both bulk and boundary perspectives
combined with symmetry arguments. We also discuss feasible experimental
platforms to realize our model. 
\end{abstract}
\maketitle
\section{Introduction} 
Two-dimensional (2D) massless
Dirac states have attracted tremendous attention in condensed matter
physics and material science since the successful realization of graphene
\cite{Novoselov05nature,Castro09rmp,LiuCC11prl,Malko12prl,Young15prl}.
Dirac states in graphene exhibit particular transport properties
such as anomalous quantum Hall effect \cite{ZhangYB05nat}, minimum
conductivity \cite{Ando02jpsj,Tworzydl06prl}, and Klein tunneling
\cite{Katsnelson06natphys,Stander09prl}. Dirac points in graphene are pinned to the corners of the hexagonal Brillouin zone (BZ) by $C_{3v}$ group symmetry of the honeycomb lattice. They can only be slightly shifted by applying external strain \cite{Pereira09prb}. As the large momentum separation of the Dirac point ensures the well-defined valley degrees of freedom, graphene provides many interesting applications such as valley filters \cite{Rycerz07NP} and valleytronics \cite{YaoW07prl,YaoW08prb}.

If we are able to manipulate Dirac cones, such as their locations and shape, we may change the properties of the system significantly. For instance, the merging of two Dirac points can transform the system from a semimetal to a trivial insulator \cite{Montambaux09prb,Feilhauer15prb} and deformed Dirac cones show anisotropic transport \cite{Pereira09prb}. Therefore, searching for alternative 2D platforms that host Dirac states with conveniently tunable properties, such as the synthetic honeycomb lattice \cite{Feilhauer15prb,Wunsch08njp,Tarruell12nature,Bellec13prl,Real20prl}, is of fundamental interest. 

\begin{figure}
\includegraphics[width=1\linewidth]{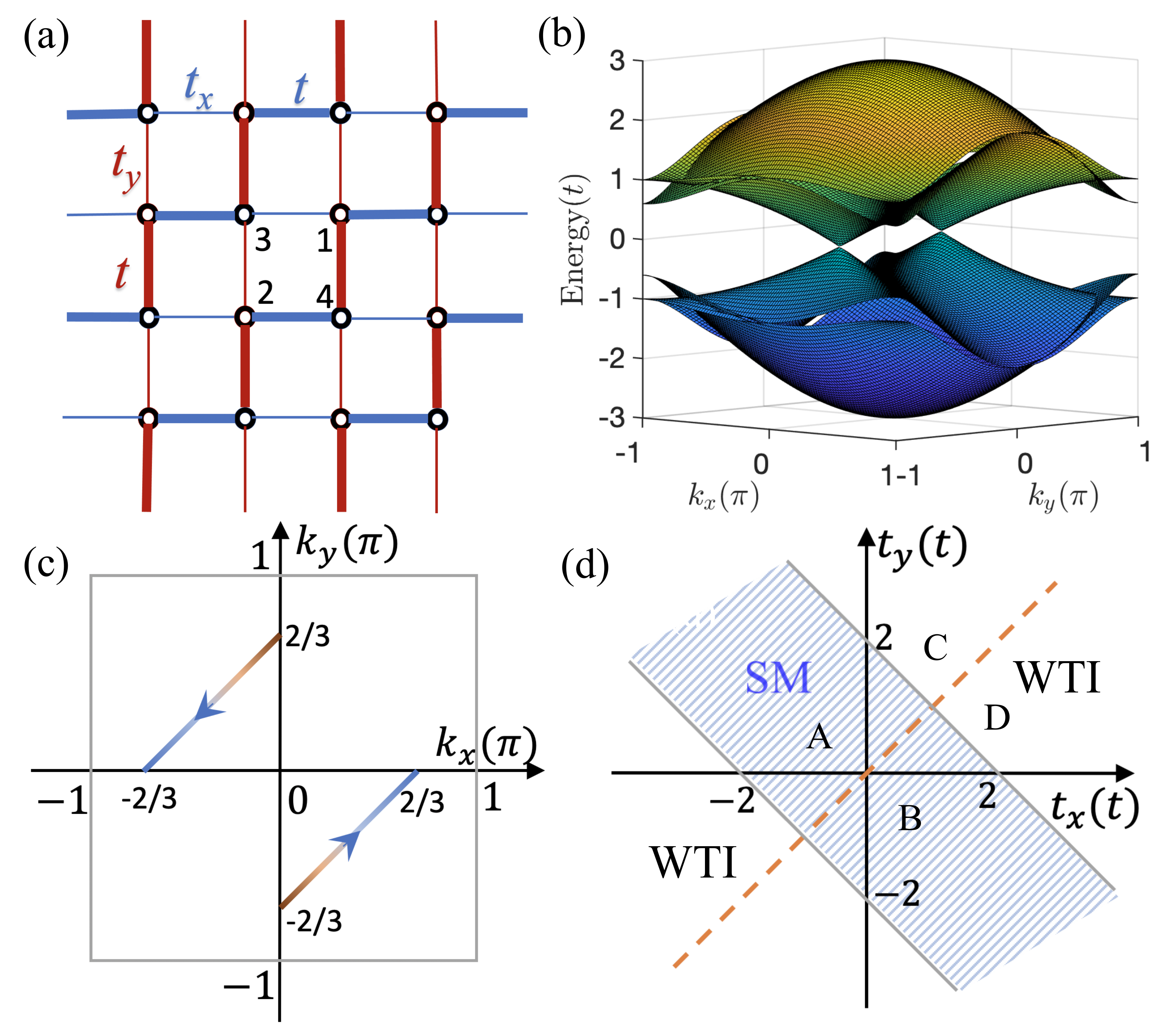}
\caption{(a) Schematic of the inclined 2D SSH lattice. Blue (red) thick and
thin bonds mark alternately dimerized hopping strengths in $x(y)$-direction.
(b) Band structure of the inclined 2D SSH model with a pair of Dirac
points. Here, we take $t_{x}=0.2t$ and $t_{y}=0.8t$. (c) Moving
pattern of the pair of Dirac points corresponding to the condition
in Eq.~\eqref{eq:Dirac movement}. (d) Phase diagram of the inclined
2D SSH model in the $(t_{x},t_{y})$ parameter space. The shadowed
region (excluding the orange dashed line) represents the semimetal
(SM) with a pair of Dirac points. The orange dashed line at $t_{x}=t_{y}$
corresponds to the nodal-line semimetal. Other regions denote the
weak topological insulator (WTI) regime. The four points $A,B,C$,
and $D$ label the representative phase points that we refer to in
the text. \label{fig:lattice}}
\end{figure}

In this work, we propose to realize 2D Dirac states with tunable properties on a Su-Schrieffer-Heeger (SSH) square lattice. Recently,
research activities related to the generalization of the SSH model
\cite{SSH79prl} to 2D have attracted broad interest \cite{LiuF17prl,Benalcazar17Science,BBH17prb}
and sparked the fast-expanding field of higher-order topological insulators
\cite{SongZD17prl,Schindler18SA,Schindler18NP,Imhof18np,Serra-Garcia18nature,XieBY19prl,ChenXD19prl,Peterson18nature,LiL18prb,Jeon22prb,ZhangRX20prl,LiCA20prl,WeiQ21prl,LiCA20prb,ZhangSB20prr,LiCA21prl}.
In our proposal, we consider an inclined 2D SSH model with alternately dimerized
patterns \textcolor{black}{{[}Fig.\ \ref{fig:lattice}}(a){]}. 
Different from previous 2D SSH models \cite{LiuF17prl,Benalcazar17Science,BBH17prb}, such an inclined 2D SSH model hosts massless Dirac states in the
BZ within a broad parameter range \textcolor{black}{{[}Figs.\ \ref{fig:lattice}}(b,d){]}. We show that the Dirac points are protected by a space-time inversion
symmetry. Moreover, we find that the locations of Dirac points are highly tunable
by hopping parameters \textcolor{black}{{[}Fig.\ \ref{fig:lattice}}(c){]}.
In contrast to artificial graphene, the merging of two Dirac points in our
model experiences a particular topological phase transition resulting
in topological phases, i.e., such as weak topological insulator or nodal-line semimetal \textcolor{black}{{[}Fig.\ \ref{fig:lattice}}(d){]}. 
We demonstrate the topological origin of these  phases by
employing topological invariants, boundary signatures,
and symmetry arguments. We also discuss how to realize our model experimentally
based on synthetic quantum materials.

The remainder of this paper is organized as follows. Section II introduces the inclined 2D SSH model. Section III presents the tunable Dirac states on the inclined 2D SSH model. Section IV discusses symmetry protection on these Dirac states. Section V considers the topological phase transition induced by merging of Dirac points. Section VI exhibits the special anisotropic projection properties of Dirac states in our model. Section VII concludes our results with a discussion. Technical details are discussed in four appendices. 

\section{Inclined two-dimensional Su-Schrieffer-Heeger model} 
We consider a particular 2D SSH model, as shown in \textcolor{black}{Fig.\ \ref{fig:lattice}}(a),
where the weak (thin) bonds and strong (thick) bonds are alternately
dimerized along the two adjacent parallel lattice rows ($x$-direction)
or columns ($y$-direction). We call it the inclined 2D SSH model.
There are four orbital degrees of freedom in each unit cell (labeled
as $1-4$). We consider spinless fermions, for clarity. The effective
Bloch Hamiltonian describing the inclined 2D SSH model in reciprocal
space reads
\begin{equation}
H({\bf k})=\left(\begin{array}{cc}
0 & q({\bf k})\\
q^{\dagger}({\bf k}) & 0
\end{array}\right),\label{eq:Hamiltonian}
\end{equation}
\begin{equation}
q({\bf k})\equiv\left(\begin{array}{cc}
t_{x}+te^{ik_{x}} & t+t_{y}e^{ik_{y}}\\
t_{y}+te^{-ik_{y}} & t+t_{x}e^{-ik_{x}}
\end{array}\right),
\end{equation}
where ${\bf k}=(k_{x},k_{y})$ is the 2D wave-vector; $t$ and $t_{x/y}$
are the staggered hopping amplitudes along $x/y$-directions. Without
loss of generality, we set the lattice constant to be unity and assume
$t>0$ hereafter. The basis is $(\Psi_{{\bf k}1},\Psi_{{\bf k}2},\Psi_{{\bf k}3},\Psi_{{\bf k}4})$
of Bloch states constructed on the four sites of the unit cell. The
Hamiltonian in Eq.~\eqref{eq:Hamiltonian} respects chiral (sublattice)
symmetry, as indicated by its block off-diagonal form. Explicitly,
chiral symmetry yields $\mathcal{C}H({\bf k})\mathcal{C}^{-1}=-H({\bf k})$
with chiral-symmetry operator $\mathcal{C}=\tau_{3}\otimes\sigma_{0}$,
where $\tau$ and $\sigma$ are Pauli matrices for different orbital
degrees of freedom in the unit cell. The energy bands of Eq.~\eqref{eq:Hamiltonian}
are obtained as 
\begin{equation}
E_{\eta}^{\pm}({\bf k})=\pm\sqrt{\xi_{\eta}^{2}({\bf k})+\zeta_{\eta}^{2}({\bf k})}=\pm|\varepsilon_{\eta}({\bf k})|,\label{eq:energy spectrum}
\end{equation}
where $\xi_{\eta}({\bf k})\equiv(t+t_{x})\cos\frac{k_{x}}{2}+\eta(t+t_{y})\cos\frac{k_{y}}{2}$,
$\zeta_{\eta}({\bf k})\equiv(t-t_{x})\sin\frac{k_{x}}{2}-\eta(t-t_{y})\sin\frac{k_{y}}{2}$,
and $\varepsilon_{\eta}({\bf k})\equiv\xi_{\eta}({\bf k})+i\zeta_{\eta}({\bf k})$
with $\eta=\pm1$ (see Appendix S1). Note that even though Eq.~\eqref{eq:Hamiltonian}
cannot be expressed in terms of anticommutating Dirac matrices only,
its energy spectrum still has the corresponding form, i.e., a square
root of the summation of some squared variables.

\section{Tunable Dirac states on square lattices} 
A pair of Dirac points appear in the BZ of the inclined 2D SSH model,
as shown in \textcolor{black}{Fig.\ \ref{fig:lattice}}(b). Interestingly,
the Dirac points are not pinned to high-symmetry points but are 
tunable by parameter modulations. To elucidate this property, it is
instructive to obtain their locations analytically. Due to the presence
of chiral symmetry, the conduction and valence bands touch at zero
energy. Thus, the existence of Dirac points yields the conditions
$\xi_{\eta}({\bf k})=\zeta_{\eta}({\bf k})=0$. Solving these condition
equations, we find a pair of Dirac points located at ${\bf K}_{\pm}\equiv\pm(K_{x},-K_{y})$,
where $K_{x/y}$ are given by
\begin{equation}
K_{x/y}=2\arccos\sqrt{\frac{(t+t_{y/x})^{2}(2t-t_{x}-t_{y})}{4t(t^{2}-t_{x}t_{y})}}.\label{eq:Diracposition}
\end{equation}
From Eq.\ \eqref{eq:Diracposition}, we find that a physical solution
(with real $K_{x/y}$ that corresponds to the presence of Dirac points)
only holds when $|t_{x}+t_{y}|<2t$ and $t_{x}\neq t_{y}$. The full
phase diagram of the inclined 2D SSH model is illustrated in Fig.\ \ref{fig:lattice}(d)
and will be discussed in more detail later.

Clearly, our model exhibits two Dirac points whose locations are 
tunable. To show this feature more explicitly, we consider a simple
parametrization with $t_{x}=s\in[0,t]$, $t_{y}=t-s$, and $t=1$.
Then, we find the relation
\begin{equation}
K_{x}+K_{y}=2\pi/3.\label{eq:Dirac movement}
\end{equation}
As a result, the Dirac points move along a line segment when we vary
the parameter $s$, as shown in \textcolor{black}{Fig.\ \ref{fig:lattice}}(c).
Note that no symmetries are broken as we move around Dirac points
by variation of $t_{x}$ and $t_{y}$. Moreover, the effective Fermi
velocity around the Dirac points in our model can also be manipulated
by parameter modulations (see Appendix S2).

\section{Space-time inversion symmetry protection on Dirac states}
In the unperturbed case with chiral symmetry, the Dirac points are
topologically described by a quantized charge $Q_{{\bf K_{\pm}}}=\frac{1}{2\pi i}\oint_{\ell}d{\bf k}\cdot\mathrm{Tr}\left[q^{-1}({\bf k})\nabla_{{\bf k}}q({\bf k})\right]$ \cite{Schnyder08prb,Schnyder11prb,Heikkila11jetp}, where
the loop $\ell$ is chosen such that it encircles a single Dirac point
${\bf K}_{\pm}$. The two Dirac points in the BZ have opposite topological
charges $Q_{{\bf K}_{\pm}}=\pm1$. They annihilate each other when
they meet in ${\bf k}$-space. In a more general sense, the stability
of the Dirac points in our model is protected by a space-time inversion
symmetry which is composed of an inversion symmetry and time-reversal
symmetry \cite{FangC15prb,Kim17prl}. Note that due to the alternate
dimerization along adjacent two lattice rows or columns, the usual
inversion symmetry is broken. If we, however, perform a glide operation
(half-unit translation) followed by an inversion operation, then the
system goes back to itself. We term this symmetry as ``glide-inversion''
symmetry. Explicitly, the glide-inversion symmetry requires
\begin{equation}
\mathcal{G}_{I}^{x,y}({\bf k})H({\bf k})[\mathcal{G}_{I}^{x,y}({\bf k})]^{-1}=H(-{\bf k}),
\end{equation}
where $\mathcal{G}_{I}^{x,y}({\bf k})=I\times g_{x/y}$, $I=\tau_{0}\otimes\sigma_{1}$
is the conventional inversion operator, $g_{x}=\tau_{1}\otimes\left(\begin{array}{cc}
e^{ik_{x}} & 0\\
0 & 1
\end{array}\right)$ and $g_{y}=\tau_{1}\otimes\left(\begin{array}{cc}
0 & e^{ik_{y}}\\
1 & 0
\end{array}\right)$ are half-unit translations along $x$- and $y$-directions, respectively.
Note that this glide-inversion symmetry can be equivalently viewed
as inversion symmetry with the inversion center shifted to the bond
center of each unit cell \textcolor{black}{{[}see Fig.\ \ref{fig:lattice}}(a){]},
we still keep the term ``glide-inversion'' to indicate its ${\bf k}$-dependence
clearly. In addition, the system respects spinless time-reversal symmetry,
i.e., $\mathcal{T}H({\bf k})\mathcal{T}^{-1}=H(-{\bf k})$, where
the time-reversal symmetry operator is given by the complex conjugation
$\mathcal{T}=\mathcal{K}$. Thus, the space-time inversion operator
can be written as $\mathcal{S}=\mathcal{G}_{I}^{x,y}({\bf k})\mathcal{T}=\tau_{0}\otimes\sigma_{1}\times g_{x/y}\mathcal{K}$.
It is a local operation in ${\bf k}$-space,
\begin{equation}
\mathcal{S}H({\bf k})\mathcal{S}^{-1}=H({\bf k}),\ \ \mathcal{S}^{2}=1.
\end{equation}
Under the constraint of $\mathcal{S}$, the Berry curvature is zero
at every point in the BZ except at the Dirac points \cite{FangC15prb,Kim17prl}.
Hence, the quantized $\pi$ Berry phase around a Dirac point protects
its stability. Indeed, if we add a staggered onsite potential as a
perturbation, say $\Delta\tau_{3}\sigma_{0}$ with $\Delta$ indicating
its strength, to break the glide-inversion symmetry, the Dirac points
are removed and a bulk gap opens (see Appendix S3). If we, however,
consider another type of staggered onsite potential $\Delta\tau_{0}\sigma_{3}$,
which breaks chiral symmetry, while it respects glide-inversion symmetry,
then the Dirac points remain intact. Therefore,
the $\pi$ Berry phase is the main topological quantity that protects
the Dirac points in our model since the chiral topological charge
$Q_{{\bf K}_{\pm}}$ needs chiral symmetry to be well-defined. 

\begin{figure}
\includegraphics[width=1\linewidth]{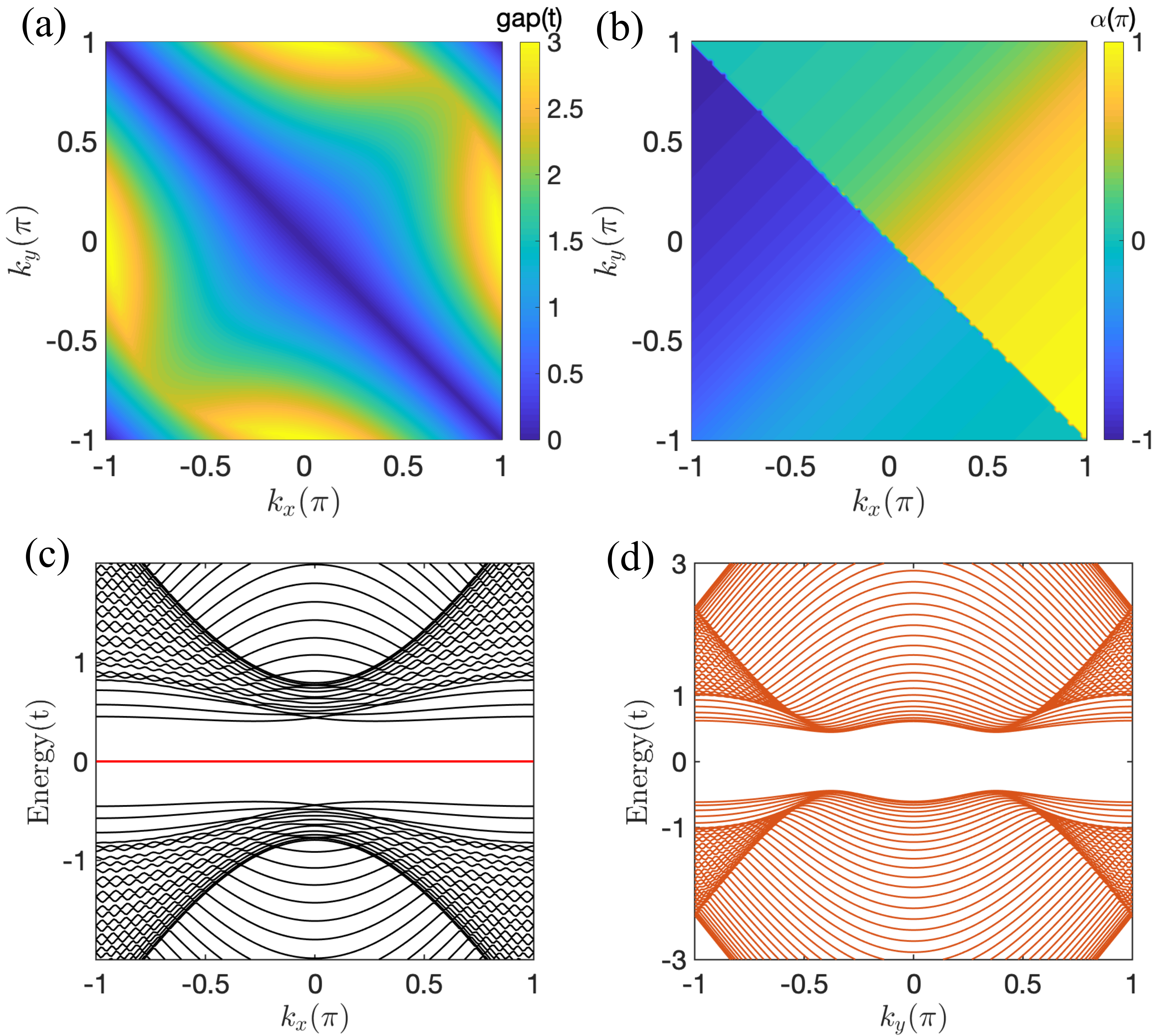}
\caption{(a) Energy gap in the whole BZ for $t_{x}=t_{y}=0.5$. A gapless nodal
line appears at the line $k_{x}+k_{y}=0$. (b) Angle $\alpha\equiv\mathrm{arg}[\varepsilon_{\eta=+1}({\bf k})]$
in the BZ for $t_{x}=t_{y}=0.5$. There is a branch cut at the line
$k_{x}+k_{y}=0$. (c) Energy spectrum of a ribbon along $x$-direction
with width $W_{y}=20$. Notice the flat band at zero energy. (d) Energy
spectrum of the ribbons along $y$-direction with width $W_{x}=20$.
Here we choose $t_{x}=1.2t,t_{y}=1.8t$ such that $w_{x}=0,w_{y}=1$.
\label{fig:nodalline}}
\end{figure}

\section{Topological phase transitions with merging of Dirac points}
The mergence of two Dirac points undergoes a
topological phase transition in our model. The semimetal phase with
a pair of Dirac points locates in the shadowed region $|t_{x}+t_{y}|<2t$
and $t_{x}\neq t_{y}$ of \textcolor{black}{Fig.\ \ref{fig:lattice}}(d).
The topological phase transition after merging a pair of Dirac points
transforms the semimetal phase to either a weak topological insulator
or a nodal-line semimetal. Thus, our inclined 2D SSH model actually
possesses three different topological phases, as shown in the phase
diagram \textcolor{black}{in Fig.\ \ref{fig:lattice}}(d). Let us
first focus on the nodal-line semimetal phase under the specific condition
$t_{x}=t_{y}$ {[}\textcolor{black}{Fig.\ \ref{fig:lattice}}(d)
and \textcolor{black}{Fig.\ \ref{fig:nodalline}}(a){]}. Consider
a representative phase point $A$ (or $B$) in the semimetal phase
with a pair of Dirac points {[}\textcolor{black}{see Fig.\ \ref{fig:lattice}}(d){]}.
As it moves towards the orange line, the Dirac points merge and we
observe that the system exhibits a gapless nodal line at
\begin{equation}
k_{x}+k_{y}=0,\ \mathrm{if}\ t_{x}=t_{y}\neq t,
\end{equation}
The appearance of a gapless nodal line is a direct consequence of
an accidental mirror symmetry. Under the condition $t_x=t_y$,
the system has mirror symmetry along the direction $x+y=0$. In momentum
space, we thus transform the Hamiltonian as $MH(k_{x},k_{y})M^{-1} =H(-k_{y},-k_{x})$ 
where the mirror operator is given by $M=\left(\begin{array}{cc}
\sigma_x & 0\\
0 & \sigma_0
\end{array}\right)$. 
Note that the Hamiltonian $H({\bf k})$ commutes with the mirror operator
$M$ along the nodal-line $k_{x}+k_{y}=0$. Therefore, we can label
the eigen states of the Hamiltonian $H({\bf k})$ by the eigen states $|\pm\rangle$
of the mirror operator $M$ as $H({\bf k})|\pm\rangle =\pm E|\pm\rangle$.
We further note that the mirror operator commutes with the chiral symmetry
operator, i.e., $[\mathcal{C},M]=0$. Therefore, we can show that
$\mathcal{C}|+\rangle$ is also an eigenstate of $M$ with eigenvalue
$+1$. Moreover, $\mathcal{C}|+\rangle$ is an eigenstate of $H({\bf k})$
with energy $+E$. Actually, chiral symmetry maps the state $|+\rangle$
with energy $+E$ to the state $\mathcal{C}|+\rangle$ with energy $-E$.
This implies that those states are degenerate at $E=0$.

The nodal-line semimetal phase is protected by a topological invariant
$\delta$ as we describe below. Let us define an angle $\alpha$
as $\alpha({\bf k})\equiv\mathrm{arg}[\varepsilon_{\eta=+1}({\bf k})]$
where $\varepsilon_{\eta}({\bf k})$ is defined below Eq.~\eqref{eq:energy spectrum}.
In the nodal-line semimetal phase, $\varepsilon_{\eta=+1}({\bf k})=(t+t_{x})(\cos\frac{k_{x}}{2}-\cos\frac{k_{y}}{2})+i(t-t_{x})(\sin\frac{k_{x}}{2}+\sin\frac{k_{y}}{2}).$ Figure \textcolor{black}{\ref{fig:nodalline}}(a) plots the band gap in the whole BZ, which clearly shows a nodal-line along $k_x=-k_y$.  
Figure \textcolor{black}{\ref{fig:nodalline}}(b) shows the angle
$\alpha$ in the BZ in the nodal-line semimetal phase. The branch
cut at $k_{x}+k_{y}=0$ separates the BZ into two equal sections.
Consider two mirror-symmetric wave-vectors ${\bf k}_{1}$ and ${\bf k}_{2}$ with respect to the line $k_x+k_y=0$. Consequently, the relation $\alpha({\bf k}_1)=\alpha({\bf k}_2)+ (2N+1)\pi$ with $N$ integer holds. 
Thus, the topological invariant $\delta$ can be defined as $\delta\equiv[\alpha({\bf k}_{1})-\alpha({\bf k}_{2})]/\pi\  \mathrm{mod}\ 2$. This topological invariant $\delta$ is protected by mirror symmetry and is not affected by the gauge degrees of freedom of $\varepsilon_{\eta}({\bf k})$. In the
case of Figs. \textcolor{black}{\ref{fig:nodalline}}(a) and \textcolor{black}{\ref{fig:nodalline}}(b),
it is found that $\delta=1$.

Again, we consider a representative phase point $A$ (or $B$) in
the semimetal phase in \textcolor{black}{Fig.\ \ref{fig:lattice}}(d).
As it moves parallel to the orange line, two Dirac points merge at
the phase boundary $|t_{x}+t_{y}|=2t$. Similar to the case of graphene,
the energy spectrum stays linear along one direction while it becomes
parabolic along another direction at the critical merging points \cite{Montambaux09prb}.
Interestingly, this topological phase transition gives rise to a weak topological
insulator rather than a trivial insulator (see the representative
phase points $C$ and D). The weak topological insulators are located
in the region $|t_{x}+t_{y}|>2t$ and $t_{x}\neq t_{y}$. They are
described by two winding numbers $(w_{x},w_{y})$ with one of them
being one and the other one being zero. The winding number is defined
as $w_{x/y}=\frac{1}{2\pi i}\int_{0}^{2\pi}dk_{x/y}\mathrm{Tr}[q^{-1}({\bf k})\partial_{k_{x/y}}q({\bf k})]$
for arbitrary $k_{y/x}\in[0,2\pi]$. Actually, this anisotropic topological insulating
phase can be further divided into two subphases: (i) $w_{x}=1,w_{y}=0$
($t_{x}>t_{y}$ and $|t_{x}+t_{y}|>2t$) and (ii) $w_{x}=0,w_{y}=1$
($t_{x}<t_{y}$ and $|t_{x}+t_{y}|>2t$).  When $w_{x}=1,w_{y}=0$ ($w_{x}=0,w_{y}=1$),
the system is nontrivial along $x$($y$)-direction and trivial along
$y(x)$-direction. Correspondingly, a totally
flat edge band exists in the gap of the energy spectrum along $x$($y$)-direction
for the subphase (i) (subphase (ii)) {[}see, for instance, \textcolor{black}{Figs.\ \ref{fig:nodalline}}(c,d){]}.
Notably, neither the topologically trivial phase with $w_{x}=w_{y}=0$
nor the topologically nontrivial phase with $w_{x}=w_{y}=1$ appear
in the inclined 2D SSH model.

\begin{figure}
\includegraphics[width=1\linewidth]{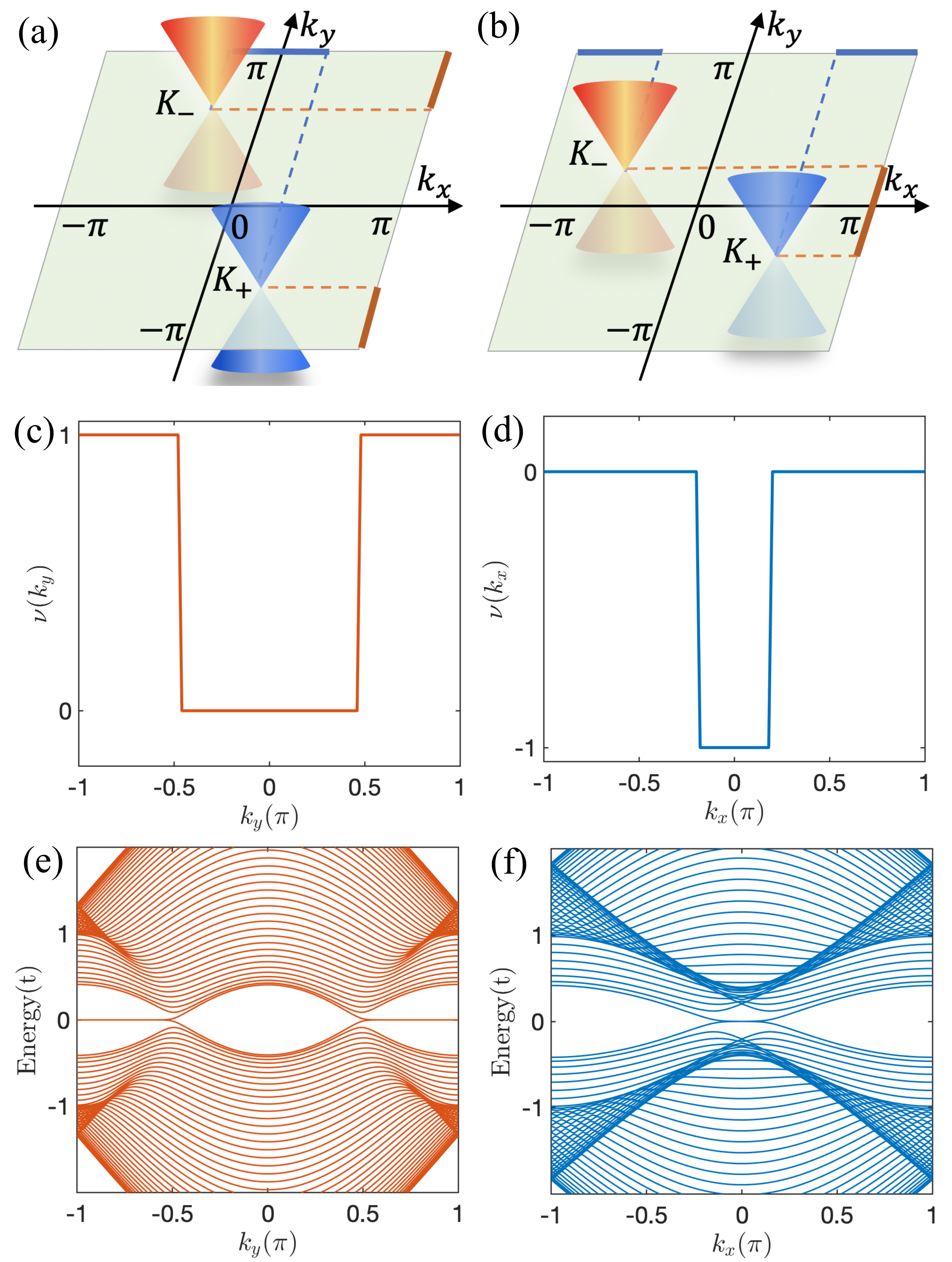}

\caption{Anisotropic projection of the two Dirac points. (a) Schematic of anisotropic
projection of bulk spectrum with two Dirac points (${\bf K}_{\pm}$)
to $x$- and $y$-directions under the condition $t_{x}<t_{y}$. The
dashed lines represent the projection direction. The thick solid lines
(blue and red) at the boundary of the BZ indicate the nontrivial regions
with flat edge bands. (b) Alternative projection similar to the case
in (a) but under the condition $t_{x}>t_{y}$. (c) Winding number
$\nu(k_{y})$ as a function of $k_{y}$. (d) Winding number $\nu(k_{x})$
as a function of $k_{x}$. (e) Energy spectrum of a ribbon along $y$-direction
with width $W_{x}=20$ corresponding to the case in panel (c). Notice
the flat band at zero energy. (f) Energy spectrum of the ribbons along
$x$-direction with width $W_{y}=20$ corresponding to the the case
in panel (d). In panels (c,d,e,f), the other parameters are $t_{x}=0.3t$
and $t_{y}=0.7t$, i.e., corresponding to the case in (a). \label{fig:anisotropic property}}
\end{figure}

\section{Anisotropic projection of Dirac points}
The projection of two Dirac points to different directions shows two
patterns as illustrated in \textcolor{black}{Figs.\ \ref{fig:anisotropic property}(a)
and \ref{fig:anisotropic property}(b).} \textcolor{black}{These two
}patterns\textcolor{black}{{} can be described by the representative
phase points $A$ and $B$ in Fig.\ \ref{fig:lattice}(d), respectively.
To switch between the two possible patterns, a phase point has to
cross the gapless nodal-line semimetal phase. For simplicity, let
us focus on the pattern in Fig.\ \ref{fig:anisotropic property}(a).
When projecting the system along $x(y)$-direction, }the same region
in the BZ can be viewed as topologically nontrivial or trivial depending
on relative values of $t_{x}$ and $t_{y}$ {[}see, for instance,
Fig.\ \ref{fig:anisotropic property}(a) for $t_{x}<t_{y}${]}. This
anisotropic nature in our model suggests two independent indexes $\nu(k_{x})$
and $\nu(k_{y})$. Explicitly, they are given by
\begin{alignat}{1}
\nu(k_{x/y}) & =\frac{1}{2\pi}\int_{0}^{2\pi}dk_{y/x}\frac{t^{2}e^{\mp ik_{y/x}}-t_{y/x}^{2}e^{\pm ik_{y/x}}}{R({\bf k})},
\end{alignat}
where $R({\bf k})\equiv(te^{ik_{x}/2}+t_{x}e^{-ik_{x}/2})^{2}-(te^{-ik_{y}/2}+t_{y}e^{ik_{y}/2})^{2}$.
In essence, $\nu(k_{x/y})$ is a winding number of the reduced one-dimensional
system at specific wave-number $k_{x/y}$ \cite{LiC172D,LiCA19jpcm}.
\textcolor{black}{In Fig.\ \ref{fig:anisotropic property}(c), the
middle region is trivial ($\nu=0$) while the two outer regions are
nontrivial ($\nu=1$). It is interesting to see that this pattern
is in stark contrast to that of Fig.\ \ref{fig:anisotropic property}(d),
i.e., $\nu=-1$ in the middle region and $\nu=0$ otherwise. This
difference comes from the anisotropic nature of the alternating dimerization
pattern in our system. }The nontrivial winding number indicates the
existence of flat edge bands at open boundaries \cite{Ryu02prl,Delplace11prb,LiC172D,LiCA19jpcm}.
Evidently, the regions of flat bands in \textcolor{black}{Figs.\ \ref{fig:anisotropic property}(e)
and \ref{fig:anisotropic property}(f) agree with the topological
nontrivial regions in Figs.\ \ref{fig:anisotropic property}(c) and
\ref{fig:anisotropic property}(d), respectively.}

\section{Discussion and conclusion}
To realize the inclined 2D SSH model experimentally, it needs a square lattice geometry
(four sites in a unit cell) and controllable nearest-neighbor couplings.
Required techniques for designing such a lattice structure have been
developed in synthetic quantum materials such as photonic and acoustic
crystals \cite{WangZ09nature,XieBY19prl,ChenXD19prl,Serra-Garcia18nature,Ni19nm},
electric circuits \cite{Imhof18np}, and waveguides \cite{Peterson18nature,Cerjan21prl}.
For instance, to realize our model in a photonic waveguide system,
waveguides can be arranged to a square lattice with four waveguides
contained in each unit cell and the alternately dimerized couplings
between neighboring waveguides can be modulated by their spacings
\cite{Cerjan21prl,Hassan19NPho}. Notably, our inclined 2D SSH model does not
require delicate manipulations of external flux, differently from recent reports to realize Dirac states
on square lattices \cite{Xue21arXiv,QiuC21arxiv,Shao21prl}. These
proposals require external $\pi$ fluxes, and the Dirac points are
pinned at boundaries of the BZ. 

Our model has a richer topological phase space accompanied with corresponding
topological phase transitions as compared to artificial graphene \cite{Tarruell12nature}. In a special limit, our model reduces to the brick-type lattice model mimicking artificial graphene \cite{Wakabayashi99prb} (see Appendix S4). Moreover, the
realization of our model on a square lattice is simpler than the
Mielke checkerboard model \cite{Mielke92JPA} since it only involves
nearest-neighbor hopping terms.   We notice that similar 2D WTIs are discussed in related works \cite{LiL18prb, Jeon22prb,Yang22nano} but no nodal-line phase is mentioned therein. 
Interestingly, our model may provide a platform to realize the toric-code insulator \cite{Tam22prb}.

In conclusion, we have proposed an inclined 2D SSH model on a square lattice
to realize highly tunable Dirac states. We have found that the locations
of Dirac points are not pinned to any high-symmetry points or lines
in the BZ but movable by parameter modifications. The mergence of
two Dirac points leads to a topological phase transition, which converts
the system from a semimetal phase with a pair of Dirac points to either
a weak topological insulator or a nodal-line semimetal. We expect
that our model can be realized in different metamaterial platforms.

\section{Acknowledgement}
This work was supported by the DFG (SPP1666 and SFB1170 ``ToCoTronics''),
the Würzburg-Dresden Cluster of Excellence ct.qmat, EXC2147, Project-id
390858490, the Elitenetzwerk Bayern Graduate School on ``Topological
Insulators'', and the High Tech Agenda Bayern.

\appendix
\numberwithin{equation}{section}
\global\long\def\thesection{S\arabic{section}}
\global\long\def\thesubsection{\Alph{subsection}}
%\begin{widetext}
\begin{center}
%\textbf{\large{}Appendices}{\large{} }
\par\end{center}{\large \par}
%\tableofcontents
\section{Properties of the inclined 2D SSH model}
In this appendix, we present the energy spectrum, Dirac points, and phase diagram of the inclined 2D SSH model. 
\subsection{Energy spectrum}
Let us calculate the spectrum of Hamiltonian Eq.~\eqref{eq:Hamiltonian} in the main text. To
do so, we utilize the general properties of chiral symmetry. In
the proper space, chiral symmetry can be expressed as $\tau_{3}H({\bf k})\tau_{3}=-H({\bf k})$.
If we consider the eigen equation 
\begin{alignat}{1}
H({\bf k})\Psi_{n} & =E_{n}({\bf k})\Psi_{n},\ \Psi_{n}=\frac{1}{\sqrt{2}}\left(\begin{array}{c}
\psi_{n}^{A}\\
\psi_{n}^{B}
\end{array}\right),
\end{alignat}
where $\psi_{n}^{A}$ and $\psi_{n}^{B}$ are the states referring
to $A$ and $B$ sublattices, respectively, then, due to chiral
symmetry, there is another state $\mathcal{C}\Psi_{n}$ satisfying the eigen equation $H({\bf k})[\mathcal{C}\Psi_{n}]=-E_{n}({\bf k})[\mathcal{C}\Psi_{n}].$
Squaring the Hamiltonian $H({\bf k})$, this yields 
\begin{alignat}{1}
H^{2}({\bf k})\Psi_{n} & =E_{n}^{2}({\bf k})\Psi_{n}.
\end{alignat}
Explicitly, we effectively decouple the equation as 
\begin{alignat}{1}
h_{A}({\bf k})\psi_{n}^{A} & =E_{n}^{2}\psi_{n}^{A},\ \ h_{A}({\bf k})\equiv q({\bf k})q^{\dagger}({\bf k}),\\
h_{B}({\bf k})\psi_{n}^{B} & =E_{n}^{2}\psi_{n}^{B},\ \ h_{B}({\bf k})\equiv q^{\dagger}({\bf k})q({\bf k}).
\end{alignat}
Note that $q({\bf k})$ is not necessarily Hermitian, while the two
defined operators $h_{A}({\bf k})$ and $h_{B}({\bf k})$ are Hermitian. 

For our model, we obtain
\begin{alignat}{1}
h_{A}({\bf k}) & =h_{0}({\bf k})\sigma_{0}+h_{1}({\bf k})\left(\begin{array}{cc}
0 & e^{i(p_x+p_y)}\\
e^{-i(p_x+p_y)} & 0
\end{array}\right),\\
h_{0}({\bf k}) & =t^{2}+2t_{x}t\cos k_{x}+t_{x}^{2}+t^{2}+2t_{y}t\cos k_{y}+t_{y}^{2},\\
h_{1}({\bf k}) & =2(t+t_{x})(t+t_{y})\cos p_x\cos p_y\nonumber\\
&-2(t-t_{x})(t-t_{y})\sin p_x \sin p_y,
\end{alignat}
where $\ p_{x}\equiv\frac{k_{x}}{2},p_{y}\equiv\frac{k_{y}}{2}$.
Thus, the energy of the system is 
\begin{alignat}{1}
E_{\eta}^{\pm}({\bf k}) & =\pm\sqrt{h_{0}({\bf k})+\eta h_{1}({\bf k})}\equiv\pm|\varepsilon_{\eta}({\bf k})|.
\end{alignat}
We have defined that 
\begin{alignat}{1}
\varepsilon_{\eta}({\bf k}) & \equiv te^{ip_{x}}+t_{x}e^{-ip_{x}}+\eta[te^{-ip_{y}}+t_{y}e^{ip_{y}}].
\end{alignat}
Then, the eigen states for $h_{A}({\bf k})$ can be simply written
as 
\begin{alignat}{1}
\psi_{\eta}^{A} & =\frac{1}{\sqrt{2}}\left(\begin{array}{c}
1\\
\eta e^{-i(p_{x}+p_{y})}
\end{array}\right).
\end{alignat}
Similar procedures can be applied to the operator $h_{B}({\bf k})$ to obtain the eigen states $\psi_{\eta}^{B}$.
Therefore, the total wave function for $H({\bf k})$ is 
\begin{alignat}{1}
\Psi_{\eta}^{\pm} & =\frac{1}{\sqrt{2}}\left(\begin{array}{c}
\psi_{\eta}^{A}\\
\pm\psi_{\eta}^{B}
\end{array}\right)=\frac{1}{2}\left(\begin{array}{c}
1\\
\eta e^{-i(p_{x}+p_{y})}\\
\pm1\\
\pm\eta e^{-i(p_{y}-p_{x})}
\end{array}\right).
\end{alignat}

\subsection{Dirac points}
Due to chiral symmetry, the conduction bands and valence bands touch
at $E=0$. For the general case $t_{x}\neq t_{y}$, it requires the
constraint $|\varepsilon_{\eta}({\bf k})|=0$ to have Dirac points,
which implies the conditions 

\begin{alignat}{1}
(t-t_{x})\sin p_{x}+\eta(t-t_{y})\sin p_{y}=0,\\
(t+t_{x})\cos p_{x}-\eta(t+t_{y})\cos p_{y}=0.
\end{alignat}
Further simplifying these equations, we can identify the locations of
the Dirac point at ${\bf K}_{\pm}=\pm(K_{x},-K_{y})$ with 

\begin{alignat}{1}
K_{x} & =2\arccos\sqrt{\frac{(t+t_{y})^{2}(2t-t_{x}-t_{y})}{4t(t^{2}-t_{x}t_{y})}},\\
K_{y} & =2\arccos\sqrt{\frac{(t+t_{x})^{2}(2t-t_{x}-t_{y})}{4t(t^{2}-t_{x}t_{y})}}.
\end{alignat}

The Dirac points are characterized by topological charges $Q_{{\bf K}_{\pm}}$
that can be calculated with the formula
\begin{alignat}{1}
Q_{{\bf K_{\pm}}}=\frac{1}{2\pi i}\oint_{\ell}d{\bf k}\cdot\mathrm{Tr}\left[q^{-1}({\bf k})\nabla_{{\bf k}}q({\bf k})\right],
\end{alignat}
where the loop $\ell$ is chosen such that it encircles a single Dirac
point ${\bf K}_{\pm}$. 
Then, we obtain
\begin{alignat}{1}
Q_{{\bf K_{\pm}}} & =\frac{1}{2\pi}\oint_{\ell}dk_{x}\frac{e^{ik_{x}}t^{2}-t_{x}^{2}e^{-ik_{x}}}{R(k_{x},k_{y})}\nonumber\\
&+\frac{1}{2\pi}\oint_{\ell}dk_{y}\frac{e^{-ik_{y}}t^{2}-t_{y}^{2}e^{ik_{y}}}{R(k_{x},k_{y})}.
\end{alignat}
We find that for any loop enclosing a single Dirac point, we get a
nonzero topological charge $Q_{{\bf K_{\pm}}}=\pm1$.

\subsection{Phase diagram}
Our inclined 2D SSH model experiences three different phases: (i)
semimetal phase with Dirac points; (ii) nodal-line semimetal phase;
(iii) weak topological insulator phase. Let us further specify the
three phases below. 

Phase (i): To obtain Dirac points, we consider the ranges of $\cos^{2}p_{x}$
and $\cos^{2}p_{y}$ as
\begin{alignat}{1}
0<\frac{(t+t_{y})^{2}(2t-t_{x}-t_{y})}{4t(t^{2}-t_{x}t_{y})}<1 & ,\\
0<\frac{(t+t_{x})^{2}(2t-t_{x}-t_{y})}{4t(t^{2}-t_{x}t_{y})}<1 & .
\end{alignat}
The conditions $\cos^{2}p_{x}>0$ and $\cos^{2}p_{y}>0$ lead to the
inequalities 
\begin{alignat}{1}
(a):\begin{cases}
2t-t_{x}-t_{y}>0,\\
t^{2}-t_{x}t_{y}>0;
\end{cases} & b):\begin{cases}
2t-t_{x}-t_{y}<0,\\
t^{2}-t_{x}t_{y}<0.
\end{cases}
\end{alignat}
For the first case (a), if we further assume $t>|(t_{x}+t_{y})/2|$,
then $t^{2}>[(t_{x}+t_{y})/2]^{2}$.
From the famous inequality that $[(a+b)/2]^2\geq ab$ for any $a,b$,
the condition $t^{2}>t_{x}t_{y}$ is always satisfied. Otherwise, if $t<|(t_{x}+t_{y})/2|$, the second
condition $t^{2}>t_{x}t_{y}$ is not always satisfied. For the second
case (b), since $t>0$, we need $(t_{x}+t_{y})/2>0$. Therefore, the
two conditions are not always compatible. 

We further need 
\begin{alignat}{1}
\frac{(t+t_{y})^{2}(2t-t_{x}-t_{y})}{4t(t^{2}-t_{x}t_{y})}<1.
\end{alignat}
It can be proven that in the regime $2t>|t_{x}+t_{y}|$ we have $t^{2}-t_{x}t_{y}>0$
and $2t-(t_{x}+t_{y})>0.$ The above condition is equivalent to  $(t-t_{y})^{2}(2t+t_{x}+t_{y}) >0$, which is also naturally true in the region $2t>|t_{x}+t_{y}|$. Therefore, the semi-metallic phase with Dirac points appear for 
\begin{equation}
2t>|t_{x}+t_{y}|.
\end{equation}

Phase (ii): From the unique form of energy spectrum, it is clear that
the system has a gapless nodal line at
\begin{equation}
k_{x}+k_{y}=0,\ \mathrm{if}\ t_{x}=t_{y}\neq t.
\end{equation}
The appearance of a gapless nodal line is due to an accidental mirror
symmetry. 

Phase (iii): The weak topological insulator phase appears in the region
$|t_{x}+t_{y}|>2t$ and $t_{x}\neq t_{y}$. The phase boundary between
weak topological insulator and semimetal phase is located at $|t_{x}+t_{y}|=2t$. 

\section{Anisotropic Fermi velocity at Dirac points}
\begin{figure}
\includegraphics[width=1\linewidth]{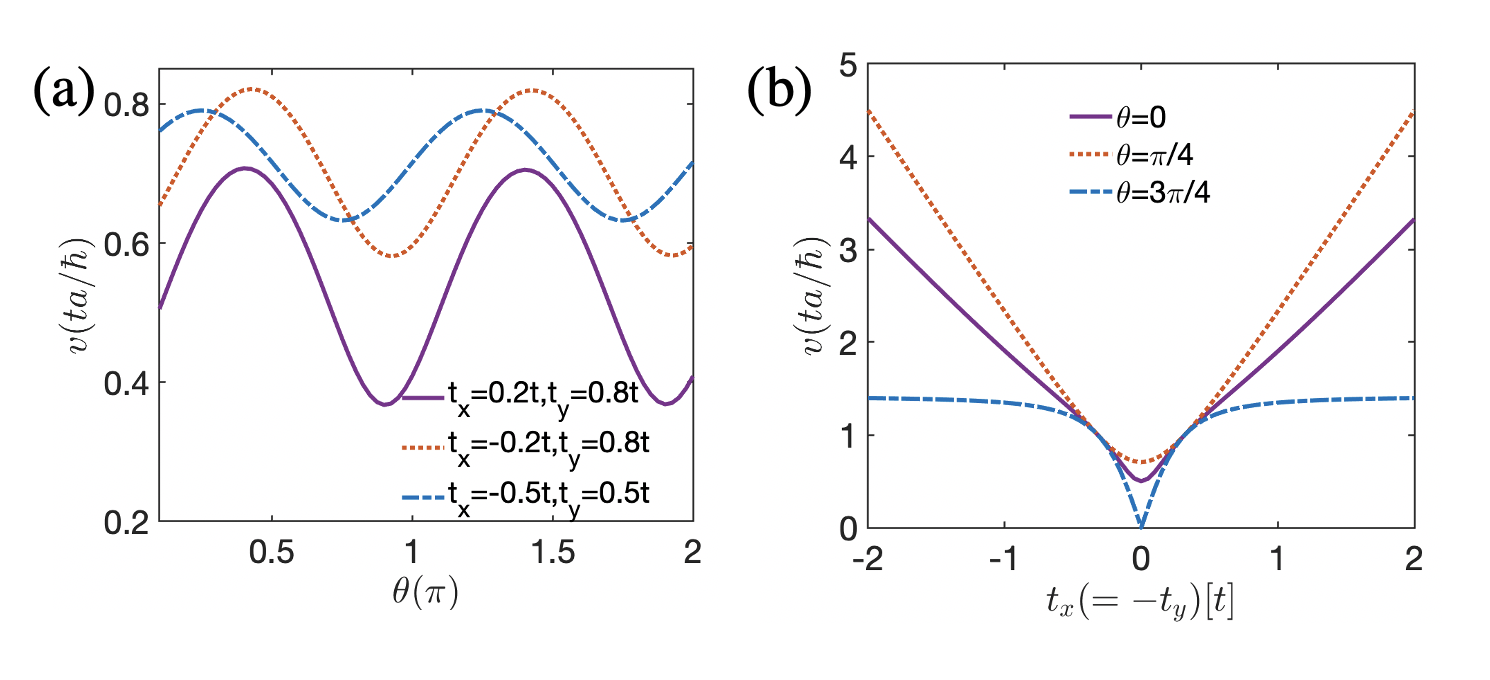}
\caption{(a) Anisotropy of the Fermi velocity at the Dirac point ${\bf K_{+}}$
as a function of the angle $\theta$ for different $t_{x}$ and $t_{y}$.
Here, $\theta$ is defined as an angle between the wave-vector ${\bf k}$
and $+\hat{k}_{x}$ axis. (b) Anisotropy of the Fermi velocity at
the Dirac point ${\bf K_{+}}$ as a function of $t_{x}(=-t_{y})$
for different angles $\theta$. \label{fig:Fermivelocity}}
\end{figure}

The effective Fermi velocity $v$ of the Dirac points in our model
can also be manipulated by parameter modulations. The effective Fermi
velocity plays a crucial role in characterizing the transport properties
of Dirac states. Let us define an angle $\theta$ between the wave-vector
${\bf k}$ and the $+\hat{k}_{x}$ axis. As shown in \textcolor{black}{Fig.\ \ref{fig:Fermivelocity}}(a),
the anisotropic Fermi velocity shows sinusoidal behavior with respect
to the angle $\theta$, consistent with elliptical Dirac cones in
the two-band effective model around ${\bf K}_{\pm}$. As tuning $t_{x}$
and $t_{y}$, the Fermi velocity in different directions $\theta$
may change substantially {[}\textcolor{black}{Fig.\ \ref{fig:Fermivelocity}}(b){]}.
Interestingly, the velocities along different directions coincide
when $t_{x}=-t_{y}\simeq\pm0.3t$, which indicates that the Dirac
cones become isotropic.

\section{Gap the Dirac points by perturbations}
\begin{figure}
\includegraphics[width=1\linewidth]{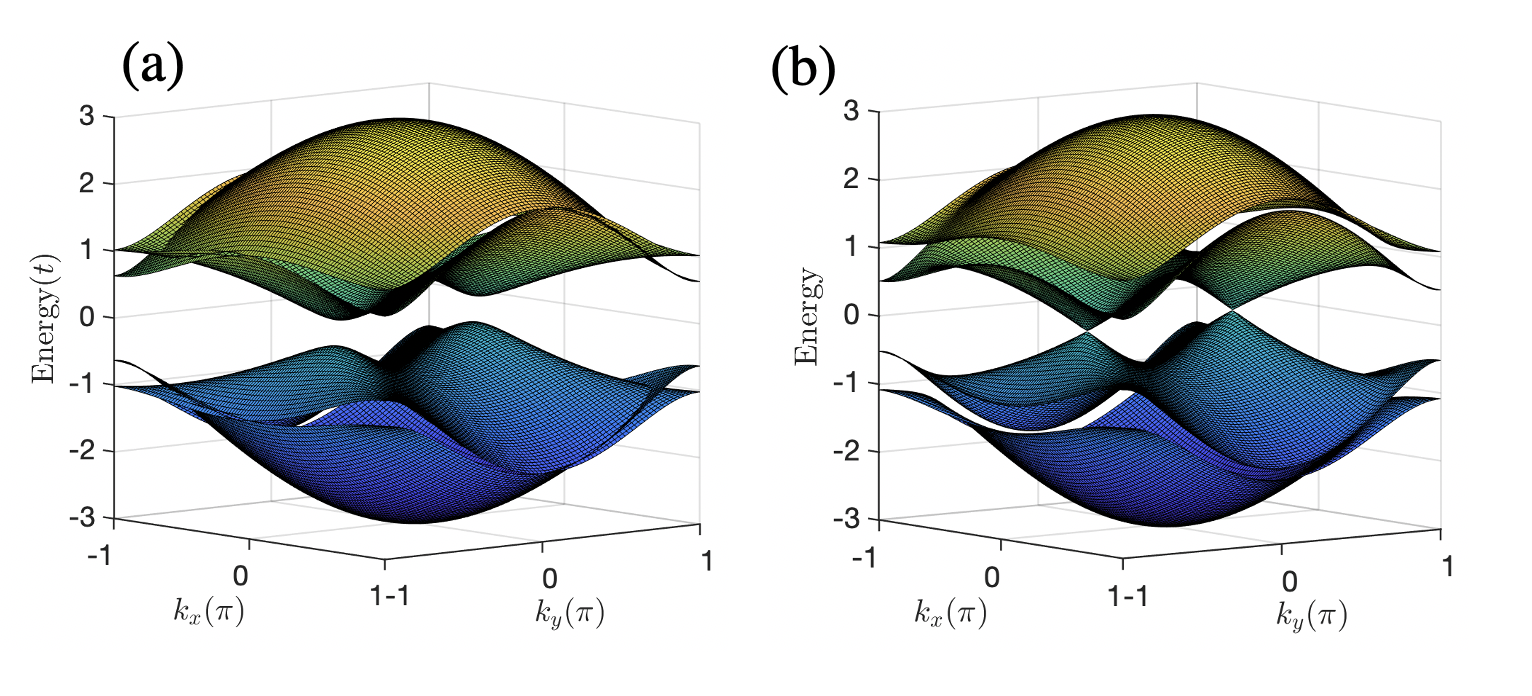}

\caption{(a) Band structure of the inclined 2D SSH model with a pair of Dirac
points being gapped by $\Delta\tau_{3}\sigma_{0}$ with $\Delta=0.1$.
(b) Band structure of the inclined 2D SSH model with a pair of Dirac
points under the perturbation term $\Delta\tau_{0}\sigma_{3}$ with
$\Delta=0.1$. Other parameters are $t_{x}=0.2t$ and $t_{y}=0.8t$.
\label{fig:perturbations}}
\end{figure}

As we discussed in the main text, the stability of Dirac
points is protected by space-time inversion symmetry. Indeed, if we
add a staggered onsite potential, say $\Delta\tau_{3}\sigma_{0}$
with $\Delta$ indicating its strength, to break the glide-inversion
symmetry, the Dirac points are removed and a bulk gap opens, as shown
in \textcolor{black}{Fig.\ \ref{fig:perturbations}}(a). If we, however,
consider another type of staggered onsite potential $\Delta\tau_{0}\sigma_{3}$,
which breaks chiral symmetry, while it respects the glide-inversion
symmetry, then the Dirac points remain intact [{Fig.\ \ref{fig:perturbations}}(b)]. 

\section{Graphene limit}
\begin{figure}
\includegraphics[width=1\linewidth]{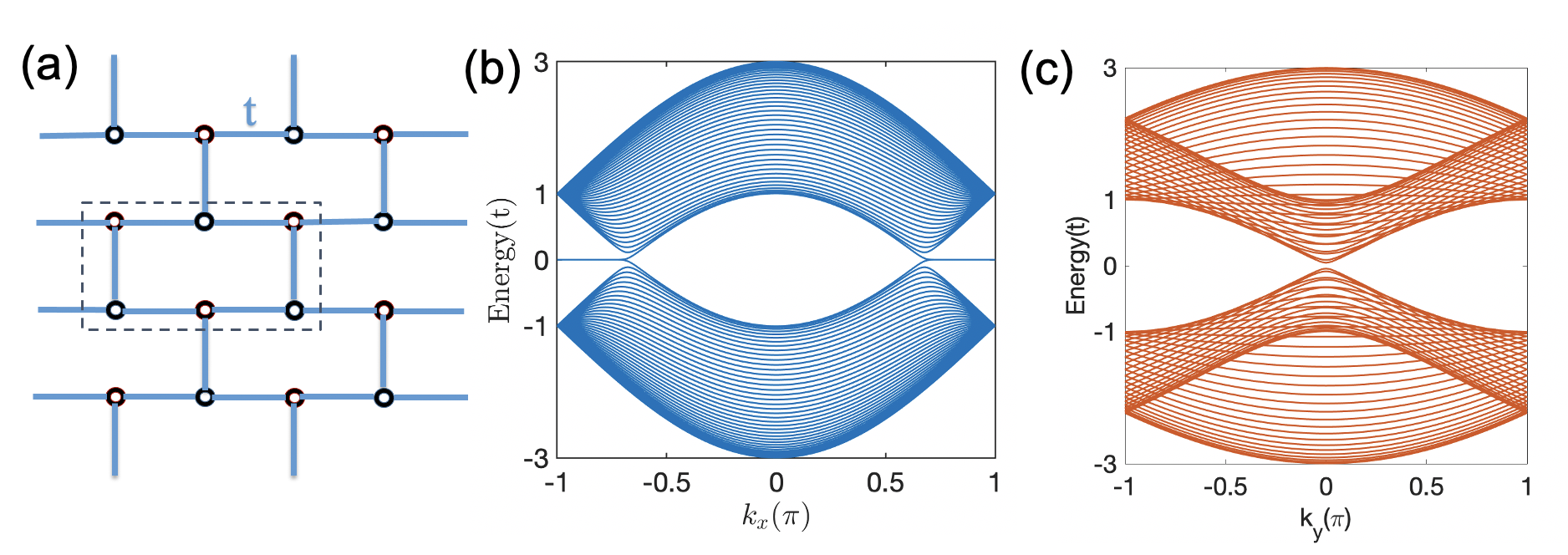}

\caption{(a) Special limit of the inclined 2D SSH model with $t_{x}=t$ and
$t_{y}=0$. (b) Spectrum of ribbon with finite width $W_{y}=20$ in
the $y$-direction corresponds to panel (a). (c) Spectrum of ribbon
with finite width $W_{x}=20$ in the $x$-direction corresponds to
panel (a). The spectrum in (b) and (c) are totally the same as the
spectrum of graphene ribbon with zigzag and armchair edges, respectively.
\label{fig:special limits}}
\end{figure}

Here, we show that our inclined 2D SSH model can reduce
to graphene (brick type lattice model) in a special limit. For instance,
it is equivalent to a square lattice version of graphene when $t_{x}=t,t_{y}=0$.
In \textcolor{black}{Fig.\ \ref{fig:special limits}}(a), the six
sites in the dashed square resemble the hexagon in graphene. Therefore,
when projecting the system to $x$-direction, the Dirac points are
located at $(\pm\frac{2\pi}{3},0)$ {[}see \textcolor{black}{Fig.\ \ref{fig:special limits}}(b){]}.
For the spectrum along perpendicular direction, there are no flat
edge bands {[}see \textcolor{black}{Fig.\ \ref{fig:special limits}}(c){]}.
Similar results are obtained if we consider the limit $t_{x}=0,t_{y}=t$. 

%\end{widetext}
%


\begin{thebibliography}{87}%
\bibliographystyle{apsrev4-1}
\makeatletter
\providecommand \@ifxundefined [1]{%
 \@ifx{#1\undefined}
}%
\providecommand \@ifnum [1]{%
 \ifnum #1\expandafter \@firstoftwo
 \else \expandafter \@secondoftwo
 \fi
}%
\providecommand \@ifx [1]{%
 \ifx #1\expandafter \@firstoftwo
 \else \expandafter \@secondoftwo
 \fi
}%
\providecommand \natexlab [1]{#1}%
\providecommand \enquote  [1]{``#1''}%
\providecommand \bibnamefont  [1]{#1}%
\providecommand \bibfnamefont [1]{#1}%
\providecommand \citenamefont [1]{#1}%
\providecommand \href@noop [0]{\@secondoftwo}%
\providecommand \href [0]{\begingroup \@sanitize@url \@href}%
\providecommand \@href[1]{\@@startlink{#1}\@@href}%
\providecommand \@@href[1]{\endgroup#1\@@endlink}%
\providecommand \@sanitize@url [0]{\catcode `\\12\catcode `\$12\catcode
  `\&12\catcode `\#12\catcode `\^12\catcode `\_12\catcode `\%12\relax}%
\providecommand \@@startlink[1]{}%
\providecommand \@@endlink[0]{}%
\providecommand \url  [0]{\begingroup\@sanitize@url \@url }%
\providecommand \@url [1]{\endgroup\@href {#1}{\urlprefix }}%
\providecommand \urlprefix  [0]{URL }%
\providecommand \Eprint [0]{\href }%
\providecommand \doibase [0]{http://dx.doi.org/}%
\providecommand \selectlanguage [0]{\@gobble}%
\providecommand \bibinfo  [0]{\@secondoftwo}%
\providecommand \bibfield  [0]{\@secondoftwo}%
\providecommand \translation [1]{[#1]}%
\providecommand \BibitemOpen [0]{}%
\providecommand \bibitemStop [0]{}%
\providecommand \bibitemNoStop [0]{.\EOS\space}%
\providecommand \EOS [0]{\spacefactor3000\relax}%
\providecommand \BibitemShut  [1]{\csname bibitem#1\endcsname}%
\let\auto@bib@innerbib\@empty
%</preamble>
\bibitem [{\citenamefont {Novoselov}\ \emph {et~al.}(2005)\citenamefont
  {Novoselov}, \citenamefont {Geim}, \citenamefont {Morozov}, \citenamefont
  {Jiang}, \citenamefont {Katsnelson}, \citenamefont {Grigorieva},
  \citenamefont {Dubonos},\ and\ \citenamefont {Firsov}}]{Novoselov05nature}%
  \BibitemOpen
  \bibfield  {author} {\bibinfo {author} {\bibfnamefont {K.~S.}\ \bibnamefont
  {Novoselov}}, \bibinfo {author} {\bibfnamefont {A.~K.}\ \bibnamefont {Geim}},
  \bibinfo {author} {\bibfnamefont {S.~V.}\ \bibnamefont {Morozov}}, \bibinfo
  {author} {\bibfnamefont {D.}~\bibnamefont {Jiang}}, \bibinfo {author}
  {\bibfnamefont {M.~I.}\ \bibnamefont {Katsnelson}}, \bibinfo {author}
  {\bibfnamefont {I.~V.}\ \bibnamefont {Grigorieva}}, \bibinfo {author}
  {\bibfnamefont {S.~V.}\ \bibnamefont {Dubonos}}, \ and\ \bibinfo {author}
  {\bibfnamefont {A.~A.}\ \bibnamefont {Firsov}},\ }\textit{Two-dimensional gas of massless Dirac fermions in graphene},  \href
  {http://dx.doi.org/10.1038/nature04233} {\bibfield  {journal} {\bibinfo
  {journal} {Nature}\ }\textbf {\bibinfo {volume} {438}},\ \bibinfo {pages}
  {197} (\bibinfo {year} {2005})}\BibitemShut {NoStop}%
\bibitem [{\citenamefont {Castro~Neto}\ \emph {et~al.}(2009)\citenamefont
  {Castro~Neto}, \citenamefont {Guinea}, \citenamefont {Peres}, \citenamefont
  {Novoselov},\ and\ \citenamefont {Geim}}]{Castro09rmp}%
  \BibitemOpen
  \bibfield  {author} {\bibinfo {author} {\bibfnamefont {A.~H.}\ \bibnamefont
  {Castro~Neto}}, \bibinfo {author} {\bibfnamefont {F.}~\bibnamefont {Guinea}},
  \bibinfo {author} {\bibfnamefont {N.~M.~R.}\ \bibnamefont {Peres}}, \bibinfo
  {author} {\bibfnamefont {K.~S.}\ \bibnamefont {Novoselov}}, \ and\ \bibinfo
  {author} {\bibfnamefont {A.~K.}\ \bibnamefont {Geim}},\ }\textit{The electronic properties of graphene},   \href {\doibase
  10.1103/RevModPhys.81.109} {\bibfield  {journal} {\bibinfo  {journal} {Rev.
  Mod. Phys.}\ }\textbf {\bibinfo {volume} {81}},\ \bibinfo {pages} {109}
  (\bibinfo {year} {2009})}\BibitemShut {NoStop}%
\bibitem [{\citenamefont {Liu}\ \emph {et~al.}(2011)\citenamefont {Liu},
  \citenamefont {Feng},\ and\ \citenamefont {Yao}}]{LiuCC11prl}%
  \BibitemOpen
  \bibfield  {author} {\bibinfo {author} {\bibfnamefont {C.-C.}\ \bibnamefont
  {Liu}}, \bibinfo {author} {\bibfnamefont {W.}~\bibnamefont {Feng}}, \ and\
  \bibinfo {author} {\bibfnamefont {Y.}~\bibnamefont {Yao}},\ }\textit{Quantum Spin Hall Effect in Silicene and Two-Dimensional Germanium},   \href {\doibase
  10.1103/PhysRevLett.107.076802} {\bibfield  {journal} {\bibinfo  {journal}
  {Phys. Rev. Lett.}\ }\textbf {\bibinfo {volume} {107}},\ \bibinfo {pages}
  {076802} (\bibinfo {year} {2011})}\BibitemShut {NoStop}%
\bibitem [{\citenamefont {Malko}\ \emph {et~al.}(2012)\citenamefont {Malko},
  \citenamefont {Neiss}, \citenamefont {Vi\~nes},\ and\ \citenamefont
  {G\"orling}}]{Malko12prl}%
  \BibitemOpen
  \bibfield  {author} {\bibinfo {author} {\bibfnamefont {D.}~\bibnamefont
  {Malko}}, \bibinfo {author} {\bibfnamefont {C.}~\bibnamefont {Neiss}},
  \bibinfo {author} {\bibfnamefont {F.}~\bibnamefont {Vi\~nes}}, \ and\
  \bibinfo {author} {\bibfnamefont {A.}~\bibnamefont {G\"orling}},\ }\textit{Competition for Graphene: Graphynes with Direction-Dependent Dirac Cones},    \href
  {\doibase 10.1103/PhysRevLett.108.086804} {\bibfield  {journal} {\bibinfo
  {journal} {Phys. Rev. Lett.}\ }\textbf {\bibinfo {volume} {108}},\ \bibinfo
  {pages} {086804} (\bibinfo {year} {2012})}\BibitemShut {NoStop}%
\bibitem [{\citenamefont {Young}\ and\ \citenamefont
  {Kane}(2015)}]{Young15prl}%
  \BibitemOpen
  \bibfield  {author} {\bibinfo {author} {\bibfnamefont {S.~M.}\ \bibnamefont
  {Young}}\ and\ \bibinfo {author} {\bibfnamefont {C.~L.}\ \bibnamefont
  {Kane}},\ }\textit{Dirac Semimetals in Two Dimensions},  \href {\doibase 10.1103/PhysRevLett.115.126803} {\bibfield
  {journal} {\bibinfo  {journal} {Phys. Rev. Lett.}\ }\textbf {\bibinfo
  {volume} {115}},\ \bibinfo {pages} {126803} (\bibinfo {year}
  {2015})}\BibitemShut {NoStop}%
\bibitem [{\citenamefont {Zhang}\ \emph {et~al.}(2005)\citenamefont {Zhang},
  \citenamefont {Tan}, \citenamefont {Stormer},\ and\ \citenamefont
  {Kim}}]{ZhangYB05nat}%
  \BibitemOpen
  \bibfield  {author} {\bibinfo {author} {\bibfnamefont {Y.}~\bibnamefont
  {Zhang}}, \bibinfo {author} {\bibfnamefont {Y.-W.}\ \bibnamefont {Tan}},
  \bibinfo {author} {\bibfnamefont {H.~L.}\ \bibnamefont {Stormer}}, \ and\
  \bibinfo {author} {\bibfnamefont {P.}~\bibnamefont {Kim}},\ }\textit{Experimental observation of the quantum Hall effect and Berry's phase in graphene},   \href
  {http://dx.doi.org/10.1038/nature04235} {\bibfield  {journal} {\bibinfo
  {journal} {Nature}\ }\textbf {\bibinfo {volume} {438}},\ \bibinfo {pages}
  {201} (\bibinfo {year} {2005})}\BibitemShut {NoStop}%
\bibitem [{\citenamefont {Ando}\ \emph {et~al.}(2002)\citenamefont {Ando},
  \citenamefont {Zheng},\ and\ \citenamefont {Suzuura}}]{Ando02jpsj}%
  \BibitemOpen
  \bibfield  {author} {\bibinfo {author} {\bibfnamefont {T.}~\bibnamefont
  {Ando}}, \bibinfo {author} {\bibfnamefont {Y.}~\bibnamefont {Zheng}}, \ and\
  \bibinfo {author} {\bibfnamefont {H.}~\bibnamefont {Suzuura}},\ }\textit{Dynamical Conductivity and Zero-Mode Anomaly in Honeycomb Lattices},   \href
  {\doibase 10.1143/JPSJ.71.1318} {\bibfield  {journal} {\bibinfo  {journal}
  {J. Phys. Soc. Jap.}\ }\textbf {\bibinfo {volume} {71}},\ \bibinfo {pages}
  {1318} (\bibinfo {year} {2002})}\BibitemShut {NoStop}%
\bibitem [{\citenamefont {Tworzyd\l{}o}\ \emph {et~al.}(2006)\citenamefont
  {Tworzyd\l{}o}, \citenamefont {Trauzettel}, \citenamefont {Titov},
  \citenamefont {Rycerz},\ and\ \citenamefont {Beenakker}}]{Tworzydl06prl}%
  \BibitemOpen
  \bibfield  {author} {\bibinfo {author} {\bibfnamefont {J.}~\bibnamefont
  {Tworzyd\l{}o}}, \bibinfo {author} {\bibfnamefont {B.}~\bibnamefont
  {Trauzettel}}, \bibinfo {author} {\bibfnamefont {M.}~\bibnamefont {Titov}},
  \bibinfo {author} {\bibfnamefont {A.}~\bibnamefont {Rycerz}}, \ and\ \bibinfo
  {author} {\bibfnamefont {C.~W.~J.}\ \bibnamefont {Beenakker}},\ }\textit{Sub-Poissonian Shot Noise in Graphene},  \href
  {\doibase 10.1103/PhysRevLett.96.246802} {\bibfield  {journal} {\bibinfo
  {journal} {Phys. Rev. Lett.}\ }\textbf {\bibinfo {volume} {96}},\ \bibinfo
  {pages} {246802} (\bibinfo {year} {2006})}\BibitemShut {NoStop}%
\bibitem [{\citenamefont {Katsnelson}\ \emph {et~al.}(2006)\citenamefont
  {Katsnelson}, \citenamefont {Novoselov},\ and\ \citenamefont
  {Geim}}]{Katsnelson06natphys}%
  \BibitemOpen
  \bibfield  {author} {\bibinfo {author} {\bibfnamefont {M.~I.}\ \bibnamefont
  {Katsnelson}}, \bibinfo {author} {\bibfnamefont {K.~S.}\ \bibnamefont
  {Novoselov}}, \ and\ \bibinfo {author} {\bibfnamefont {A.~K.}\ \bibnamefont
  {Geim}},\ }\textit{Chiral tunnelling and the Klein paradox in graphene},   \href {http://dx.doi.org/10.1038/nphys384} {\bibfield  {journal}
  {\bibinfo  {journal} {Nat Phys}\ }\textbf {\bibinfo {volume} {2}},\ \bibinfo
  {pages} {620} (\bibinfo {year} {2006})}\BibitemShut {NoStop}%
\bibitem [{\citenamefont {Stander}\ \emph {et~al.}(2009)\citenamefont
  {Stander}, \citenamefont {Huard},\ and\ \citenamefont
  {Goldhaber-Gordon}}]{Stander09prl}%
  \BibitemOpen
  \bibfield  {author} {\bibinfo {author} {\bibfnamefont {N.}~\bibnamefont
  {Stander}}, \bibinfo {author} {\bibfnamefont {B.}~\bibnamefont {Huard}}, \
  and\ \bibinfo {author} {\bibfnamefont {D.}~\bibnamefont {Goldhaber-Gordon}},\
  }\textit{Evidence for Klein Tunneling in Graphene p n Junctions},    \href {\doibase 10.1103/PhysRevLett.102.026807} {\bibfield  {journal}
  {\bibinfo  {journal} {Phys. Rev. Lett.}\ }\textbf {\bibinfo {volume} {102}},\
  \bibinfo {pages} {026807} (\bibinfo {year} {2009})}\BibitemShut {NoStop}%
\bibitem [{\citenamefont {Pereira}\ \emph {et~al.}(2009)\citenamefont
  {Pereira}, \citenamefont {Castro~Neto},\ and\ \citenamefont
  {Peres}}]{Pereira09prb}%
  \BibitemOpen
  \bibfield  {author} {\bibinfo {author} {\bibfnamefont {V.~M.}\ \bibnamefont
  {Pereira}}, \bibinfo {author} {\bibfnamefont {A.~H.}\ \bibnamefont
  {Castro~Neto}}, \ and\ \bibinfo {author} {\bibfnamefont {N.~M.~R.}\
  \bibnamefont {Peres}},\ }\textit{Tight-binding approach to uniaxial strain in graphene},  \href {\doibase 10.1103/PhysRevB.80.045401}
  {\bibfield  {journal} {\bibinfo  {journal} {Phys. Rev. B}\ }\textbf {\bibinfo
  {volume} {80}},\ \bibinfo {pages} {045401} (\bibinfo {year}
  {2009})}\BibitemShut {NoStop}%
\bibitem [{\citenamefont {Rycerz}\ \emph {et~al.}(2007)\citenamefont {Rycerz},
  \citenamefont {Tworzyd{\l}o},\ and\ \citenamefont {Beenakker}}]{Rycerz07NP}%
  \BibitemOpen
  \bibfield  {author} {\bibinfo {author} {\bibfnamefont {A.}~\bibnamefont
  {Rycerz}}, \bibinfo {author} {\bibfnamefont {J.}~\bibnamefont
  {Tworzyd{\l}o}}, \ and\ \bibinfo {author} {\bibfnamefont {C.~W.~J.}\
  \bibnamefont {Beenakker}},\ }\textit{Valley filter and valley valve in graphene},   \href {\doibase 10.1038/nphys547} {\bibfield
  {journal} {\bibinfo  {journal} {Nature Physics}\ }\textbf {\bibinfo {volume}
  {3}},\ \bibinfo {pages} {172} (\bibinfo {year} {2007})}\BibitemShut {NoStop}%
\bibitem [{\citenamefont {Xiao}\ \emph {et~al.}(2007)\citenamefont {Xiao},
  \citenamefont {Yao},\ and\ \citenamefont {Niu}}]{YaoW07prl}%
  \BibitemOpen
  \bibfield  {author} {\bibinfo {author} {\bibfnamefont {D.}~\bibnamefont
  {Xiao}}, \bibinfo {author} {\bibfnamefont {W.}~\bibnamefont {Yao}}, \ and\
  \bibinfo {author} {\bibfnamefont {Q.}~\bibnamefont {Niu}},\ }\textit{Valley-Contrasting Physics in Graphene: Magnetic Moment and Topological Transport},    \href {\doibase
  10.1103/PhysRevLett.99.236809} {\bibfield  {journal} {\bibinfo  {journal}
  {Phys. Rev. Lett.}\ }\textbf {\bibinfo {volume} {99}},\ \bibinfo {pages}
  {236809} (\bibinfo {year} {2007})}\BibitemShut {NoStop}%
\bibitem [{\citenamefont {Yao}\ \emph {et~al.}(2008)\citenamefont {Yao},
  \citenamefont {Xiao},\ and\ \citenamefont {Niu}}]{YaoW08prb}%
  \BibitemOpen
  \bibfield  {author} {\bibinfo {author} {\bibfnamefont {W.}~\bibnamefont
  {Yao}}, \bibinfo {author} {\bibfnamefont {D.}~\bibnamefont {Xiao}}, \ and\
  \bibinfo {author} {\bibfnamefont {Q.}~\bibnamefont {Niu}},\ }\textit{Valley-dependent optoelectronics from inversion symmetry breaking},    \href {\doibase
  10.1103/PhysRevB.77.235406} {\bibfield  {journal} {\bibinfo  {journal} {Phys.
  Rev. B}\ }\textbf {\bibinfo {volume} {77}},\ \bibinfo {pages} {235406}
  (\bibinfo {year} {2008})}\BibitemShut {NoStop}%
\bibitem [{\citenamefont {Montambaux}\ \emph {et~al.}(2009)\citenamefont
  {Montambaux}, \citenamefont {Pi\'echon}, \citenamefont {Fuchs},\ and\
  \citenamefont {Goerbig}}]{Montambaux09prb}%
  \BibitemOpen
  \bibfield  {author} {\bibinfo {author} {\bibfnamefont {G.}~\bibnamefont
  {Montambaux}}, \bibinfo {author} {\bibfnamefont {F.}~\bibnamefont
  {Pi\'echon}}, \bibinfo {author} {\bibfnamefont {J.-N.}\ \bibnamefont
  {Fuchs}}, \ and\ \bibinfo {author} {\bibfnamefont {M.~O.}\ \bibnamefont
  {Goerbig}},\ }\textit{Merging of Dirac points in a two-dimensional crystal},    \href {\doibase 10.1103/PhysRevB.80.153412} {\bibfield
  {journal} {\bibinfo  {journal} {Phys. Rev. B}\ }\textbf {\bibinfo {volume}
  {80}},\ \bibinfo {pages} {153412} (\bibinfo {year} {2009})}\BibitemShut
  {NoStop}%
\bibitem [{\citenamefont {Feilhauer}\ \emph {et~al.}(2015)\citenamefont
  {Feilhauer}, \citenamefont {Apel}, and\
  \citenamefont {Schweitzer}}]{Feilhauer15prb}%
  \BibitemOpen
  \bibfield  {author} {\bibinfo {author} {\bibfnamefont {J.}~\bibnamefont
  {Feilhauer}}, \bibinfo {author} {\bibfnamefont {W.}~\bibnamefont
  {Apel}}, and\ \bibinfo {author} {\bibfnamefont {L.}\ \bibnamefont
  {Schweitzer}},\ }\textit{Merging of the Dirac points in electronic artificial graphene},     \href {\doibase 10.1103/PhysRevB.92.245424} {\bibfield
  {journal} {\bibinfo  {journal} {Phys. Rev. B}\ }\textbf {\bibinfo {volume}
  {92}},\ \bibinfo {pages} {245424} (\bibinfo {year} {2015})}\BibitemShut
  {NoStop}% 
\bibitem [{\citenamefont {Wunsch}\ \emph {et~al.}(2008)\citenamefont
  {Wunsch}, \citenamefont {Guinea}, and\
  \citenamefont {Sols}}]{Wunsch08njp}%
  \BibitemOpen
  \bibfield  {author} {\bibinfo {author} {\bibfnamefont {B.}~\bibnamefont
  {Wunsch}}, \bibinfo {author} {\bibfnamefont {F.}~\bibnamefont
  {Guinea}}, and\ \bibinfo {author} {\bibfnamefont {F.}\ \bibnamefont
  {Sols}},\ }\textit{Dirac-point engineering and topological phase transitions in honeycomb optical lattices},      \href {\doibase 10.1088/1367-2630/10/10/103027} {\bibfield
  {journal} {\bibinfo  {journal} {N. J. Phys.}\ }\textbf {\bibinfo {volume}
  {10}},\ \bibinfo {pages} {103027} (\bibinfo {year} {2008})}\BibitemShut
  {NoStop}%    
\bibitem [{\citenamefont {Tarruell}\ \emph {et~al.}(2012)\citenamefont
  {Tarruell}, \citenamefont {Greif}, \citenamefont {Uehlinger}, \citenamefont
  {Jotzu},\ and\ \citenamefont {Esslinger}}]{Tarruell12nature}%
  \BibitemOpen
  \bibfield  {author} {\bibinfo {author} {\bibfnamefont {L.}~\bibnamefont
  {Tarruell}}, \bibinfo {author} {\bibfnamefont {D.}~\bibnamefont {Greif}},
  \bibinfo {author} {\bibfnamefont {T.}~\bibnamefont {Uehlinger}}, \bibinfo
  {author} {\bibfnamefont {G.}~\bibnamefont {Jotzu}}, \ and\ \bibinfo {author}
  {\bibfnamefont {T.}~\bibnamefont {Esslinger}},\ }\textit{Creating, moving and merging Dirac points with a Fermi gas in a tunable honeycomb lattice},      \href {\doibase
  10.1038/nature10871} {\bibfield  {journal} {\bibinfo  {journal} {Nature}\
  }\textbf {\bibinfo {volume} {483}},\ \bibinfo {pages} {302} (\bibinfo {year}
  {2012})}\BibitemShut {NoStop}%
\bibitem [{\citenamefont {Bellec}\ \emph {et~al.}(2013)\citenamefont {Bellec},
  \citenamefont {Kuhl}, \citenamefont {Montambaux},\ and\ \citenamefont {Mortessagne}}] {Bellec13prl}%
  \BibitemOpen
  \bibfield  {author} {\bibinfo {author} {\bibfnamefont {M.}~\bibnamefont
  {Bellec}}, \bibinfo {author} {\bibfnamefont {U.}~\bibnamefont
  {Kuhl}}, \bibinfo {author} {\bibfnamefont {G.}~\bibnamefont
  {Montambaux}},\ and\ \bibinfo {author} {\bibfnamefont {F.}\
  \bibnamefont {Mortessagne}},\ }\textit{Topological Transition of Dirac Points in a Microwave Experiment},     \href {\doibase 10.1103/PhysRevLett.110.033902} {\bibfield
  {journal} {\bibinfo  {journal} {Phys. Rev. Lett.}\ }\textbf {\bibinfo {volume}
  {110}},\ \bibinfo {pages} {033902} (\bibinfo {year} {2013})}\BibitemShut {NoStop}%  
\bibitem [{\citenamefont {Real}\ \emph {et~al.}(2020)\citenamefont {Real},
  \citenamefont {Jamadi}, \citenamefont {Milicevic},\ and\ \citenamefont {et. al}}] {Real20prl}%
  \BibitemOpen
  \bibfield  {author} {\bibinfo {author} {\bibfnamefont {B.}~\bibnamefont
  {Real}}, \bibinfo {author} {\bibfnamefont {O.}~\bibnamefont
  {Jamadi}}, \bibinfo {author} {\bibfnamefont {M.}~\bibnamefont
  {Milicevic}},\ and\ \bibinfo {author} {\bibfnamefont {et.,}\
  \bibnamefont {al.}},\ }\textit{Semi-Dirac Transport and Anisotropic Localization in Polariton Honeycomb Lattices},   \href {\doibase 10.1103/PhysRevLett.125.186601} {\bibfield
  {journal} {\bibinfo  {journal} {Phys. Rev. Lett.}\ }\textbf {\bibinfo {volume}
  {125}},\ \bibinfo {pages} {186601} (\bibinfo {year} {2020})}\BibitemShut {NoStop}%  
\bibitem [{\citenamefont {Su}\ \emph {et~al.}(1979)\citenamefont {Su},
  \citenamefont {Schrieffer},\ and\ \citenamefont {Heeger}}]{SSH79prl}%
  \BibitemOpen
  \bibfield  {author} {\bibinfo {author} {\bibfnamefont {W.~P.}\ \bibnamefont
  {Su}}, \bibinfo {author} {\bibfnamefont {J.~R.}\ \bibnamefont {Schrieffer}},
  \ and\ \bibinfo {author} {\bibfnamefont {A.~J.}\ \bibnamefont {Heeger}},\
  }\textit{Solitons in Polyacetylene}, \href {\doibase 10.1103/PhysRevLett.42.1698} {\bibfield  {journal} {\bibinfo
   {journal} {Phys. Rev. Lett.}\ }\textbf {\bibinfo {volume} {42}},\ \bibinfo
  {pages} {1698} (\bibinfo {year} {1979})}\BibitemShut {NoStop}%
\bibitem [{\citenamefont {Liu}\ and\ \citenamefont
  {Wakabayashi}(2017)}]{LiuF17prl}%
  \BibitemOpen
  \bibfield  {author} {\bibinfo {author} {\bibfnamefont {F.}~\bibnamefont
  {Liu}}\ and\ \bibinfo {author} {\bibfnamefont {K.}~\bibnamefont
  {Wakabayashi}},\ }\textit{Novel Topological Phase with a Zero Berry Curvature}, \href {\doibase 10.1103/PhysRevLett.118.076803} {\bibfield
  {journal} {\bibinfo  {journal} {Phys. Rev. Lett.}\ }\textbf {\bibinfo
  {volume} {118}},\ \bibinfo {pages} {076803} (\bibinfo {year}
  {2017})}\BibitemShut {NoStop}%
\bibitem [{\citenamefont {Benalcazar}\ \emph
  {et~al.}(2017{\natexlab{a}})\citenamefont {Benalcazar}, \citenamefont
  {Bernevig},\ and\ \citenamefont {Hughes}}]{Benalcazar17Science}%
  \BibitemOpen
  \bibfield  {author} {\bibinfo {author} {\bibfnamefont {W.~A.}\ \bibnamefont
  {Benalcazar}}, \bibinfo {author} {\bibfnamefont {B.~A.}\ \bibnamefont
  {Bernevig}}, \ and\ \bibinfo {author} {\bibfnamefont {T.~L.}\ \bibnamefont
  {Hughes}},\ }\textit{Quantized electric multipole insulators}, \href {\doibase 10.1126/science.aah6442} {\bibfield  {journal}
  {\bibinfo  {journal} {Science}\ }\textbf {\bibinfo {volume} {357}},\ \bibinfo
  {pages} {61} (\bibinfo {year} {2017}{\natexlab{a}})}\BibitemShut {NoStop}%
\bibitem [{\citenamefont {Benalcazar}\ \emph
  {et~al.}(2017{\natexlab{b}})\citenamefont {Benalcazar}, \citenamefont
  {Bernevig},\ and\ \citenamefont {Hughes}}]{BBH17prb}%
  \BibitemOpen
  \bibfield  {author} {\bibinfo {author} {\bibfnamefont {W.~A.}\ \bibnamefont
  {Benalcazar}}, \bibinfo {author} {\bibfnamefont {B.~A.}\ \bibnamefont
  {Bernevig}}, \ and\ \bibinfo {author} {\bibfnamefont {T.~L.}\ \bibnamefont
  {Hughes}},\ }\textit{Electric multipole moments, topological multipole moment pumping, and chiral hinge states in crystalline insulators},  \href {\doibase 10.1103/PhysRevB.96.245115} {\bibfield
  {journal} {\bibinfo  {journal} {Phys. Rev. B}\ }\textbf {\bibinfo {volume}
  {96}},\ \bibinfo {pages} {245115} (\bibinfo {year}
  {2017}{\natexlab{b}})}\BibitemShut {NoStop}%
\bibitem [{\citenamefont {Song}\ \emph {et~al.}(2017)\citenamefont {Song},
  \citenamefont {Fang},\ and\ \citenamefont {Fang}}]{SongZD17prl}%
  \BibitemOpen
  \bibfield  {author} {\bibinfo {author} {\bibfnamefont {Z.}~\bibnamefont
  {Song}}, \bibinfo {author} {\bibfnamefont {Z.}~\bibnamefont {Fang}}, \ and\
  \bibinfo {author} {\bibfnamefont {C.}~\bibnamefont {Fang}},\ }\textit{(d-2)-Dimensional Edge States of Rotation Symmetry Protected Topological States},  \href {\doibase
  10.1103/PhysRevLett.119.246402} {\bibfield  {journal} {\bibinfo  {journal}
  {Phys. Rev. Lett.}\ }\textbf {\bibinfo {volume} {119}},\ \bibinfo {pages}
  {246402} (\bibinfo {year} {2017})}\BibitemShut {NoStop}%
\bibitem [{\citenamefont {Schindler}\ \emph
  {et~al.}(2018{\natexlab{a}})\citenamefont {Schindler}, \citenamefont {Cook},
  \citenamefont {Vergniory}, \citenamefont {Wang}, \citenamefont {Parkin},
  \citenamefont {Bernevig},\ and\ \citenamefont {Neupert}}]{Schindler18SA}%
  \BibitemOpen
  \bibfield  {author} {\bibinfo {author} {\bibfnamefont {F.}~\bibnamefont
  {Schindler}}, \bibinfo {author} {\bibfnamefont {A.~M.}\ \bibnamefont {Cook}},
  \bibinfo {author} {\bibfnamefont {M.~G.}\ \bibnamefont {Vergniory}}, \bibinfo
  {author} {\bibfnamefont {Z.}~\bibnamefont {Wang}}, \bibinfo {author}
  {\bibfnamefont {S.~S.~P.}\ \bibnamefont {Parkin}}, \bibinfo {author}
  {\bibfnamefont {B.~A.}\ \bibnamefont {Bernevig}}, \ and\ \bibinfo {author}
  {\bibfnamefont {T.}~\bibnamefont {Neupert}},\ }\textit{Higher-order topological insulators},   \href
  {https://advances.sciencemag.org/content/4/6/eaat0346} {\bibfield  {journal}
  {\bibinfo  {journal} {Science Advances}\ }\textbf {\bibinfo {volume} {4}}
  (\bibinfo {year} {2018}{\natexlab{a}})}\BibitemShut {NoStop}%
\bibitem [{\citenamefont {Schindler}\ \emph
  {et~al.}(2018{\natexlab{b}})\citenamefont {Schindler}, \citenamefont {Wang},
  \citenamefont {Vergniory}, \citenamefont {Cook}, \citenamefont {Murani},
  \citenamefont {Sengupta}, \citenamefont {Kasumov}, \citenamefont {Deblock},
  \citenamefont {Jeon}, \citenamefont {Drozdov}, \citenamefont {Bouchiat},
  \citenamefont {Gu{\'e}ron}, \citenamefont {Yazdani}, \citenamefont
  {Bernevig},\ and\ \citenamefont {Neupert}}]{Schindler18NP}%
  \BibitemOpen
  \bibfield  {author} {\bibinfo {author} {\bibfnamefont {F.}~\bibnamefont
  {Schindler}}, \bibinfo {author} {\bibfnamefont {Z.}~\bibnamefont {Wang}},
  \bibinfo {author} {\bibfnamefont {M.~G.}\ \bibnamefont {Vergniory}}, \bibinfo
  {author} {\bibfnamefont {A.~M.}\ \bibnamefont {Cook}}, \bibinfo {author}
  {\bibfnamefont {A.}~\bibnamefont {Murani}}, \bibinfo {author} {\bibfnamefont
  {S.}~\bibnamefont {Sengupta}}, \bibinfo {author} {\bibfnamefont {A.~Y.}\
  \bibnamefont {Kasumov}}, \bibinfo {author} {\bibfnamefont {R.}~\bibnamefont
  {Deblock}}, \bibinfo {author} {\bibfnamefont {S.}~\bibnamefont {Jeon}},
  \bibinfo {author} {\bibfnamefont {I.}~\bibnamefont {Drozdov}}, \bibinfo
  {author} {\bibfnamefont {H.}~\bibnamefont {Bouchiat}}, \bibinfo {author}
  {\bibfnamefont {S.}~\bibnamefont {Gu{\'e}ron}}, \bibinfo {author}
  {\bibfnamefont {A.}~\bibnamefont {Yazdani}}, \bibinfo {author} {\bibfnamefont
  {B.~A.}\ \bibnamefont {Bernevig}}, \ and\ \bibinfo {author} {\bibfnamefont
  {T.}~\bibnamefont {Neupert}},\ }\textit{Higher-order topology in bismuth},    \href {\doibase 10.1038/s41567-018-0224-7}
  {\bibfield  {journal} {\bibinfo  {journal} {Nat. Phys.}\ }\textbf {\bibinfo
  {volume} {14}},\ \bibinfo {pages} {918} (\bibinfo {year}
  {2018}{\natexlab{b}})}\BibitemShut {NoStop}%
\bibitem [{\citenamefont {Imhof}\ \emph {et~al.}(2018)\citenamefont {Imhof},
  \citenamefont {Berger}, \citenamefont {Bayer}, \citenamefont {Brehm},
  \citenamefont {Molenkamp}, \citenamefont {Kiessling}, \citenamefont
  {Schindler}, \citenamefont {Lee}, \citenamefont {Greiter}, \citenamefont
  {Neupert},\ and\ \citenamefont {Thomale}}]{Imhof18np}%
  \BibitemOpen
  \bibfield  {author} {\bibinfo {author} {\bibfnamefont {S.}~\bibnamefont
  {Imhof}}, \bibinfo {author} {\bibfnamefont {C.}~\bibnamefont {Berger}},
  \bibinfo {author} {\bibfnamefont {F.}~\bibnamefont {Bayer}}, \bibinfo
  {author} {\bibfnamefont {J.}~\bibnamefont {Brehm}}, \bibinfo {author}
  {\bibfnamefont {L.~W.}\ \bibnamefont {Molenkamp}}, \bibinfo {author}
  {\bibfnamefont {T.}~\bibnamefont {Kiessling}}, \bibinfo {author}
  {\bibfnamefont {F.}~\bibnamefont {Schindler}}, \bibinfo {author}
  {\bibfnamefont {C.~H.}\ \bibnamefont {Lee}}, \bibinfo {author} {\bibfnamefont
  {M.}~\bibnamefont {Greiter}}, \bibinfo {author} {\bibfnamefont
  {T.}~\bibnamefont {Neupert}}, \ and\ \bibinfo {author} {\bibfnamefont
  {R.}~\bibnamefont {Thomale}},\ }\textit{Topolectrical-circuit realization of topological corner modes},   \href {\doibase 10.1038/s41567-018-0246-1}
  {\bibfield  {journal} {\bibinfo  {journal} {Nat. Phys.}\ }\textbf {\bibinfo
  {volume} {14}},\ \bibinfo {pages} {925} (\bibinfo {year} {2018})}\BibitemShut
  {NoStop}%
\bibitem [{\citenamefont {Serra-Garcia}\ \emph {et~al.}(2018)\citenamefont
  {Serra-Garcia}, \citenamefont {Peri}, \citenamefont {S{\"u}sstrunk},
  \citenamefont {Bilal}, \citenamefont {Larsen}, \citenamefont {Villanueva},\
  and\ \citenamefont {Huber}}]{Serra-Garcia18nature}%
  \BibitemOpen
  \bibfield  {author} {\bibinfo {author} {\bibfnamefont {M.}~\bibnamefont
  {Serra-Garcia}}, \bibinfo {author} {\bibfnamefont {V.}~\bibnamefont {Peri}},
  \bibinfo {author} {\bibfnamefont {R.}~\bibnamefont {S{\"u}sstrunk}}, \bibinfo
  {author} {\bibfnamefont {O.~R.}\ \bibnamefont {Bilal}}, \bibinfo {author}
  {\bibfnamefont {T.}~\bibnamefont {Larsen}}, \bibinfo {author} {\bibfnamefont
  {L.~G.}\ \bibnamefont {Villanueva}}, \ and\ \bibinfo {author} {\bibfnamefont
  {S.~D.}\ \bibnamefont {Huber}},\ }\textit{Observation of a phononic quadrupole topological insulator},    \href {\doibase 10.1038/nature25156}
  {\bibfield  {journal} {\bibinfo  {journal} {Nature}\ }\textbf {\bibinfo
  {volume} {555}},\ \bibinfo {pages} {342} (\bibinfo {year}
  {2018})}\BibitemShut {NoStop}%
\bibitem [{\citenamefont {Xie}\ \emph {et~al.}(2019)\citenamefont {Xie},
  \citenamefont {Su}, \citenamefont {Wang}, \citenamefont {Su}, \citenamefont
  {Shen}, \citenamefont {Zhan}, \citenamefont {Lu}, \citenamefont {Wang},\ and\
  \citenamefont {Chen}}]{XieBY19prl}%
  \BibitemOpen
  \bibfield  {author} {\bibinfo {author} {\bibfnamefont {B.-Y.}\ \bibnamefont
  {Xie}}, \bibinfo {author} {\bibfnamefont {G.-X.}\ \bibnamefont {Su}},
  \bibinfo {author} {\bibfnamefont {H.-F.}\ \bibnamefont {Wang}}, \bibinfo
  {author} {\bibfnamefont {H.}~\bibnamefont {Su}}, \bibinfo {author}
  {\bibfnamefont {X.-P.}\ \bibnamefont {Shen}}, \bibinfo {author}
  {\bibfnamefont {P.}~\bibnamefont {Zhan}}, \bibinfo {author} {\bibfnamefont
  {M.-H.}\ \bibnamefont {Lu}}, \bibinfo {author} {\bibfnamefont {Z.-L.}\
  \bibnamefont {Wang}}, \ and\ \bibinfo {author} {\bibfnamefont {Y.-F.}\
  \bibnamefont {Chen}},\ }\textit{Visualization of Higher-Order Topological Insulating Phases in Two-Dimensional Dielectric Photonic Crystals},   \href {\doibase 10.1103/PhysRevLett.122.233903}
  {\bibfield  {journal} {\bibinfo  {journal} {Phys. Rev. Lett.}\ }\textbf
  {\bibinfo {volume} {122}},\ \bibinfo {pages} {233903} (\bibinfo {year}
  {2019})}\BibitemShut {NoStop}%
\bibitem [{\citenamefont {Chen}\ \emph {et~al.}(2019)\citenamefont {Chen},
  \citenamefont {Deng}, \citenamefont {Shi}, \citenamefont {Zhao},
  \citenamefont {Chen},\ and\ \citenamefont {Dong}}]{ChenXD19prl}%
  \BibitemOpen
  \bibfield  {author} {\bibinfo {author} {\bibfnamefont {X.-D.}\ \bibnamefont
  {Chen}}, \bibinfo {author} {\bibfnamefont {W.-M.}\ \bibnamefont {Deng}},
  \bibinfo {author} {\bibfnamefont {F.-L.}\ \bibnamefont {Shi}}, \bibinfo
  {author} {\bibfnamefont {F.-L.}\ \bibnamefont {Zhao}}, \bibinfo {author}
  {\bibfnamefont {M.}~\bibnamefont {Chen}}, \ and\ \bibinfo {author}
  {\bibfnamefont {J.-W.}\ \bibnamefont {Dong}},\ }\textit{Direct Observation of Corner States in Second-Order Topological Photonic Crystal Slabs},    \href {\doibase
  10.1103/PhysRevLett.122.233902} {\bibfield  {journal} {\bibinfo  {journal}
  {Phys. Rev. Lett.}\ }\textbf {\bibinfo {volume} {122}},\ \bibinfo {pages}
  {233902} (\bibinfo {year} {2019})}\BibitemShut {NoStop}%
\bibitem [{\citenamefont {Peterson}\ \emph {et~al.}(2018)\citenamefont
  {Peterson}, \citenamefont {Benalcazar}, \citenamefont {Hughes},\ and\
  \citenamefont {Bahl}}]{Peterson18nature}%
  \BibitemOpen
  \bibfield  {author} {\bibinfo {author} {\bibfnamefont {C.~W.}\ \bibnamefont
  {Peterson}}, \bibinfo {author} {\bibfnamefont {W.~A.}\ \bibnamefont
  {Benalcazar}}, \bibinfo {author} {\bibfnamefont {T.~L.}\ \bibnamefont
  {Hughes}}, \ and\ \bibinfo {author} {\bibfnamefont {G.}~\bibnamefont
  {Bahl}},\ }\textit{A quantized microwave quadrupole insulator with topologically protected corner states},     \href {\doibase 10.1038/nature25777} {\bibfield  {journal}
  {\bibinfo  {journal} {Nature}\ }\textbf {\bibinfo {volume} {555}},\ \bibinfo
  {pages} {346} (\bibinfo {year} {2018})}\BibitemShut {NoStop}%
\bibitem [{\citenamefont {Li}\ \emph {et~al.}(2018)\citenamefont
  {Li}, \citenamefont {Umer}, \ and\
  \citenamefont {Gong}}]{LiL18prb}%
  \BibitemOpen
  \bibfield  {author} {\bibinfo {author} {\bibfnamefont {L.}\ \bibnamefont
  {Li}}, \bibinfo {author} {\bibfnamefont {M.}~\bibnamefont {Umer}},
  \ and\ \bibinfo {author} {\bibfnamefont {J.}\ \bibnamefont {Gong}},\ }\textit{Direct prediction of corner state configurations from edge winding numbers in two- and three-dimensional chiral-symmetric lattice systems},     \href{\doibase 10.1103/PhysRevB.98.205422} {\bibfield  {journal} {\bibinfo
  {journal} {Phys. Rev. B}\ }\textbf {\bibinfo {volume} {98}},\ \bibinfo
  {pages} {205422} (\bibinfo {year} {2018})}\BibitemShut {NoStop}%
\bibitem [{\citenamefont {Jeon}\ \emph {et~al.}(2022)\citenamefont
  {Joen},\ and\
  \citenamefont {Kim}}]{Jeon22prb}%
  \BibitemOpen
  \bibfield  {author} {\bibinfo {author} {\bibfnamefont {S.}\ \bibnamefont
  {Jeon}},\ and\ \bibinfo {author} {\bibfnamefont {Y.}\ \bibnamefont {Kim}},\ }\textit{Two-dimensional weak topological insulators in inversion-symmetric crystals},   \href
  {\doibase 10.1103/PhysRevB.105.L121101} {\bibfield  {journal} {\bibinfo
  {journal} {Phys. Rev. B}\ }\textbf {\bibinfo {volume} {105}},\ \bibinfo
  {pages} {L121101} (\bibinfo {year} {2022})}\BibitemShut {NoStop}%
\bibitem [{\citenamefont {Zhang}\ \emph {et~al.}(2020)\citenamefont {Zhang},
  \citenamefont {Wu},\ and\ \citenamefont {Das~Sarma}}]{ZhangRX20prl}%
  \BibitemOpen
  \bibfield  {author} {\bibinfo {author} {\bibfnamefont {R.-X.}\ \bibnamefont
  {Zhang}}, \bibinfo {author} {\bibfnamefont {F.}~\bibnamefont {Wu}}, \ and\
  \bibinfo {author} {\bibfnamefont {S.}~\bibnamefont {Das~Sarma}},\ }\textit{Möbius Insulator and Higher-Order Topology in 
MnBiTe},  \href
  {\doibase 10.1103/PhysRevLett.124.136407} {\bibfield  {journal} {\bibinfo
  {journal} {Phys. Rev. Lett.}\ }\textbf {\bibinfo {volume} {124}},\ \bibinfo
  {pages} {136407} (\bibinfo {year} {2020})}\BibitemShut {NoStop}%
\bibitem [{\citenamefont {Li}\ \emph {et~al.}(2020)\citenamefont {Li},
  \citenamefont {Fu}, \citenamefont {Hu}, \citenamefont {Li},\ and\
  \citenamefont {Shen}}]{LiCA20prl}%
  \BibitemOpen
  \bibfield  {author} {\bibinfo {author} {\bibfnamefont {C.-A.}\ \bibnamefont
  {Li}}, \bibinfo {author} {\bibfnamefont {B.}~\bibnamefont {Fu}}, \bibinfo
  {author} {\bibfnamefont {Z.-A.}\ \bibnamefont {Hu}}, \bibinfo {author}
  {\bibfnamefont {J.}~\bibnamefont {Li}}, \ and\ \bibinfo {author}
  {\bibfnamefont {S.-Q.}\ \bibnamefont {Shen}},\ }\textit{Topological Phase Transitions in Disordered Electric Quadrupole Insulators},   \href {\doibase
  10.1103/PhysRevLett.125.166801} {\bibfield  {journal} {\bibinfo  {journal}
  {Phys. Rev. Lett.}\ }\textbf {\bibinfo {volume} {125}},\ \bibinfo {pages}
  {166801} (\bibinfo {year} {2020})}\BibitemShut {NoStop}%
\bibitem [{\citenamefont {Wei}\ \emph {et~al.}(2021)\citenamefont {Wei},
  \citenamefont {Zhang}, \citenamefont {Deng}, \citenamefont {Lu},
  \citenamefont {Huang}, \citenamefont {Yan}, \citenamefont {Chen},
  \citenamefont {Liu},\ and\ \citenamefont {Jia}}]{WeiQ21prl}%
  \BibitemOpen
  \bibfield  {author} {\bibinfo {author} {\bibfnamefont {Q.}~\bibnamefont
  {Wei}}, \bibinfo {author} {\bibfnamefont {X.}~\bibnamefont {Zhang}}, \bibinfo
  {author} {\bibfnamefont {W.}~\bibnamefont {Deng}}, \bibinfo {author}
  {\bibfnamefont {J.}~\bibnamefont {Lu}}, \bibinfo {author} {\bibfnamefont
  {X.}~\bibnamefont {Huang}}, \bibinfo {author} {\bibfnamefont
  {M.}~\bibnamefont {Yan}}, \bibinfo {author} {\bibfnamefont {G.}~\bibnamefont
  {Chen}}, \bibinfo {author} {\bibfnamefont {Z.}~\bibnamefont {Liu}}, \ and\
  \bibinfo {author} {\bibfnamefont {S.}~\bibnamefont {Jia}},\ }\textit{3D Hinge Transport in Acoustic Higher-Order Topological Insulators},   \href {\doibase
  10.1103/PhysRevLett.127.255501} {\bibfield  {journal} {\bibinfo  {journal}
  {Phys. Rev. Lett.}\ }\textbf {\bibinfo {volume} {127}},\ \bibinfo {pages}
  {255501} (\bibinfo {year} {2021})}\BibitemShut {NoStop}%
\bibitem [{\citenamefont {Li}\ and\ \citenamefont
  {Wu}(2020)}]{LiCA20prb}%
  \BibitemOpen
  \bibfield  {author} {\bibinfo {author} {\bibfnamefont {C.-A.}\ \bibnamefont
  {Li}}\ and\ \bibinfo {author} {\bibfnamefont {S. S.}~\bibnamefont
  {Wu}},\ }\textit{Topological states in generalized electric quadrupole insulators},   \href {\doibase doi.org/10.1103/PhysRevB.101.195309} {\bibfield
   {journal} {\bibinfo  {journal} {Phys. Rev. B}\ }\textbf {\bibinfo
  {volume} {101}},\ \bibinfo {pages} {195309} (\bibinfo {year}
  {2020})}\BibitemShut {NoStop}%  
\bibitem [{\citenamefont {Zhang}\ and\ \citenamefont
  {Trauzettel}(2020)}]{ZhangSB20prr}%
  \BibitemOpen
  \bibfield  {author} {\bibinfo {author} {\bibfnamefont {S.-B.}\ \bibnamefont
  {Zhang}}\ and\ \bibinfo {author} {\bibfnamefont {B.}~\bibnamefont
  {Trauzettel}},\ }\textit{Detection of second-order topological superconductors by Josephson junctions},   \href {\doibase 10.1103/PhysRevResearch.2.012018} {\bibfield
   {journal} {\bibinfo  {journal} {Phys. Rev. Research}\ }\textbf {\bibinfo
  {volume} {2}},\ \bibinfo {pages} {012018} (\bibinfo {year}
  {2020})}\BibitemShut {NoStop}%
\bibitem [{\citenamefont {Li}\ \emph {et~al.}(2021{\natexlab{a}})\citenamefont
  {Li}, \citenamefont {Zhang}, \citenamefont {Li},\ and\ \citenamefont
  {Trauzettel}}]{LiCA21prl}%
  \BibitemOpen
  \bibfield  {author} {\bibinfo {author} {\bibfnamefont {C.-A.}\ \bibnamefont
  {Li}}, \bibinfo {author} {\bibfnamefont {S.-B.}\ \bibnamefont {Zhang}},
  \bibinfo {author} {\bibfnamefont {J.}~\bibnamefont {Li}}, \ and\ \bibinfo
  {author} {\bibfnamefont {B.}~\bibnamefont {Trauzettel}},\ }\textit{Higher-Order Fabry-Pérot Interferometer from Topological Hinge States},  \href {\doibase
  10.1103/PhysRevLett.127.026803} {\bibfield  {journal} {\bibinfo  {journal}
  {Phys. Rev. Lett.}\ }\textbf {\bibinfo {volume} {127}},\ \bibinfo {pages}
  {026803} (\bibinfo {year} {2021}{\natexlab{a}})}\BibitemShut {NoStop}%
\bibitem [{\citenamefont {Schnyder}\ \emph {et~al.}(2008)\citenamefont
  {Schnyder}, \citenamefont {Ryu}, \citenamefont {Furusaki},\ and\
  \citenamefont {Ludwig}}]{Schnyder08prb}%
  \BibitemOpen
  \bibfield  {author} {\bibinfo {author} {\bibfnamefont {A.~P.}\ \bibnamefont
  {Schnyder}}, \bibinfo {author} {\bibfnamefont {S.}~\bibnamefont {Ryu}},
  \bibinfo {author} {\bibfnamefont {A.}~\bibnamefont {Furusaki}}, \ and\
  \bibinfo {author} {\bibfnamefont {A.~W.~W.}\ \bibnamefont {Ludwig}},\ }\textit{Classification of topological insulators and superconductors in three spatial dimensions},  \href
  {\doibase 10.1103/PhysRevB.78.195125} {\bibfield  {journal} {\bibinfo
  {journal} {Phys. Rev. B}\ }\textbf {\bibinfo {volume} {78}},\ \bibinfo
  {pages} {195125} (\bibinfo {year} {2008})}\BibitemShut {NoStop}%
\bibitem [{\citenamefont {Schnyder}\ and\ \citenamefont
  {Ryu}(2011)}]{Schnyder11prb}%
  \BibitemOpen
  \bibfield  {author} {\bibinfo {author} {\bibfnamefont {A.~P.}\ \bibnamefont
  {Schnyder}}\ and\ \bibinfo {author} {\bibfnamefont {S.}~\bibnamefont {Ryu}},\
  }\textit{Topological phases and surface flat bands in superconductors without inversion symmetry},   \href {\doibase 10.1103/PhysRevB.84.060504} {\bibfield  {journal} {\bibinfo
  {journal} {Phys. Rev. B}\ }\textbf {\bibinfo {volume} {84}},\ \bibinfo
  {pages} {060504} (\bibinfo {year} {2011})}\BibitemShut {NoStop}%
\bibitem [{\citenamefont {Heikkil{\"a}}\ \emph {et~al.}(2011)\citenamefont
  {Heikkil{\"a}}, \citenamefont {Kopnin},\ and\ \citenamefont
  {Volovik}}]{Heikkila11jetp}%
  \BibitemOpen
  \bibfield  {author} {\bibinfo {author} {\bibfnamefont {T.~T.}\ \bibnamefont
  {Heikkil{\"a}}}, \bibinfo {author} {\bibfnamefont {N.~B.}\ \bibnamefont
  {Kopnin}}, \ and\ \bibinfo {author} {\bibfnamefont {G.~E.}\ \bibnamefont
  {Volovik}},\ }\textit{Flat bands in topological media},   \href {\doibase 10.1134/S0021364011150045} {\bibfield
  {journal} {\bibinfo  {journal} {JETP Letters}\ }\textbf {\bibinfo {volume}
  {94}},\ \bibinfo {pages} {233} (\bibinfo {year} {2011})}\BibitemShut
  {NoStop}%
\bibitem [{\citenamefont {Fang}\ and\ \citenamefont {Fu}(2015)}]{FangC15prb}%
  \BibitemOpen
\bibfield  {journal} {  }\bibfield  {author} {\bibinfo {author} {\bibfnamefont
  {C.}~\bibnamefont {Fang}}\ and\ \bibinfo {author} {\bibfnamefont
  {L.}~\bibnamefont {Fu}},\ }\textit{New classes of three-dimensional topological crystalline insulators: Nonsymmorphic and magnetic},   \href {\doibase 10.1103/PhysRevB.91.161105}
  {\bibfield  {journal} {\bibinfo  {journal} {Phys. Rev. B}\ }\textbf {\bibinfo
  {volume} {91}},\ \bibinfo {pages} {161105} (\bibinfo {year}
  {2015})}\BibitemShut {NoStop}%
\bibitem [{\citenamefont {Kim}\ \emph {et~al.}(2017)\citenamefont {Kim},
  \citenamefont {Baik}, \citenamefont {Jung}, \citenamefont {Sohn},
  \citenamefont {Ryu}, \citenamefont {Choi}, \citenamefont {Yang},\ and\
  \citenamefont {Kim}}]{Kim17prl}%
  \BibitemOpen
  \bibfield  {author} {\bibinfo {author} {\bibfnamefont {J.}~\bibnamefont
  {Kim}}, \bibinfo {author} {\bibfnamefont {S.~S.}\ \bibnamefont {Baik}},
  \bibinfo {author} {\bibfnamefont {S.~W.}\ \bibnamefont {Jung}}, \bibinfo
  {author} {\bibfnamefont {Y.}~\bibnamefont {Sohn}}, \bibinfo {author}
  {\bibfnamefont {S.~H.}\ \bibnamefont {Ryu}}, \bibinfo {author} {\bibfnamefont
  {H.~J.}\ \bibnamefont {Choi}}, \bibinfo {author} {\bibfnamefont {B.-J.}\
  \bibnamefont {Yang}}, \ and\ \bibinfo {author} {\bibfnamefont {K.~S.}\
  \bibnamefont {Kim}},\ }\textit{Two-Dimensional Dirac Fermions Protected by Space-Time Inversion Symmetry in Black Phosphorus},  \href {\doibase 10.1103/PhysRevLett.119.226801}
  {\bibfield  {journal} {\bibinfo  {journal} {Phys. Rev. Lett.}\ }\textbf
  {\bibinfo {volume} {119}},\ \bibinfo {pages} {226801} (\bibinfo {year}
  {2017})}\BibitemShut {NoStop}%
\bibitem [{\citenamefont {Shen}\ \emph {et~al.}(2017)\citenamefont {Shen},
  \citenamefont {Li},\ and\ \citenamefont {Niu}}]{LiC172D}%
  \BibitemOpen
  \bibfield  {author} {\bibinfo {author} {\bibfnamefont {S.-Q.}\ \bibnamefont
  {Shen}}, \bibinfo {author} {\bibfnamefont {C.-A.}\ \bibnamefont {Li}}, \ and\
  \bibinfo {author} {\bibfnamefont {Q.}~\bibnamefont {Niu}},\ } \textit{Chiral anomaly and anomalous finite-size conductivity in graphene},  \href
  {http://stacks.iop.org/2053-1583/4/i=3/a=035014} {\bibfield  {journal}
  {\bibinfo  {journal} {2D Materials}\ }\textbf {\bibinfo {volume} {4}},\
  \bibinfo {pages} {035014} (\bibinfo {year} {2017})}\BibitemShut {NoStop}%
\bibitem [{\citenamefont {Li}(2019)}]{LiCA19jpcm}%
  \BibitemOpen
  \bibfield  {author} {\bibinfo {author} {\bibfnamefont {C.-A.}\ \bibnamefont
  {Li}},\ }\textit{Pseudo chiral anomaly in zigzag graphene ribbons},  \href {\doibase 10.1088/1361-648x/ab4466} {\bibfield  {journal}
  {\bibinfo  {journal} {J. Phys.: Condens. Matter}\ }\textbf {\bibinfo {volume}
  {32}},\ \bibinfo {pages} {025301} (\bibinfo {year} {2019})}\BibitemShut
  {NoStop}%
\bibitem [{\citenamefont {Ryu}\ and\ \citenamefont
  {Hatsugai}(2002)}]{Ryu02prl}%
  \BibitemOpen
  \bibfield  {author} {\bibinfo {author} {\bibfnamefont {S.}~\bibnamefont
  {Ryu}}\ and\ \bibinfo {author} {\bibfnamefont {Y.}~\bibnamefont {Hatsugai}},\
  }\textit{Topological Origin of Zero-Energy Edge States in Particle-Hole Symmetric Systems},  \href {\doibase 10.1103/PhysRevLett.89.077002} {\bibfield  {journal}
  {\bibinfo  {journal} {Phys. Rev. Lett.}\ }\textbf {\bibinfo {volume} {89}},\
  \bibinfo {pages} {077002} (\bibinfo {year} {2002})}\BibitemShut {NoStop}%
\bibitem [{\citenamefont {Delplace}\ \emph {et~al.}(2011)\citenamefont
  {Delplace}, \citenamefont {Ullmo},\ and\ \citenamefont
  {Montambaux}}]{Delplace11prb}%
  \BibitemOpen
  \bibfield  {author} {\bibinfo {author} {\bibfnamefont {P.}~\bibnamefont
  {Delplace}}, \bibinfo {author} {\bibfnamefont {D.}~\bibnamefont {Ullmo}}, \
  and\ \bibinfo {author} {\bibfnamefont {G.}~\bibnamefont {Montambaux}},\
  }\textit{Zak phase and the existence of edge states in graphene},   \href {\doibase 10.1103/PhysRevB.84.195452} {\bibfield  {journal} {\bibinfo
  {journal} {Phys. Rev. B}\ }\textbf {\bibinfo {volume} {84}},\ \bibinfo
  {pages} {195452} (\bibinfo {year} {2011})}\BibitemShut {NoStop}%
\bibitem [{\citenamefont {Wang}\ \emph {et~al.}(2009)\citenamefont {Wang},
  \citenamefont {Chong}, \citenamefont {Joannopoulos},\ and\ \citenamefont
  {Solja{\v{c}}i{\'{c}}}}]{WangZ09nature}%
  \BibitemOpen
  \bibfield  {author} {\bibinfo {author} {\bibfnamefont {Z.}~\bibnamefont
  {Wang}}, \bibinfo {author} {\bibfnamefont {Y.}~\bibnamefont {Chong}},
  \bibinfo {author} {\bibfnamefont {J.~D.}\ \bibnamefont {Joannopoulos}}, \
  and\ \bibinfo {author} {\bibfnamefont {M.}~\bibnamefont
  {Solja{\v{c}}i{\'{c}}}},\ }\textit{Observation of unidirectional backscattering-immune topological electromagnetic states},  \href {\doibase 10.1038/nature08293} {\bibfield
  {journal} {\bibinfo  {journal} {Nature}\ }\textbf {\bibinfo {volume} {461}},\
  \bibinfo {pages} {772} (\bibinfo {year} {2009})}\BibitemShut {NoStop}%
\bibitem [{\citenamefont {Ni}\ \emph {et~al.}(2019)\citenamefont {Ni},
  \citenamefont {Weiner}, \citenamefont {Al{\`u}},\ and\ \citenamefont
  {Khanikaev}}]{Ni19nm}%
  \BibitemOpen
  \bibfield  {author} {\bibinfo {author} {\bibfnamefont {X.}~\bibnamefont
  {Ni}}, \bibinfo {author} {\bibfnamefont {M.}~\bibnamefont {Weiner}}, \bibinfo
  {author} {\bibfnamefont {A.}~\bibnamefont {Al{\`u}}}, \ and\ \bibinfo
  {author} {\bibfnamefont {A.~B.}\ \bibnamefont {Khanikaev}},\ }\textit{Observation of higher-order topological acoustic states protected by generalized chiral symmetry}, \href {\doibase
  10.1038/s41563-018-0252-9} {\bibfield  {journal} {\bibinfo  {journal} {Nat.
  Mater.}\ }\textbf {\bibinfo {volume} {18}},\ \bibinfo {pages} {113} (\bibinfo
  {year} {2019})}\BibitemShut {NoStop}%
\bibitem [{\citenamefont {Cerjan}\ \emph {et~al.}(2020)\citenamefont {Cerjan},
  \citenamefont {J\"urgensen}, \citenamefont {Benalcazar}, \citenamefont
  {Mukherjee},\ and\ \citenamefont {Rechtsman}}]{Cerjan21prl}%
  \BibitemOpen
  \bibfield  {author} {\bibinfo {author} {\bibfnamefont {A.}~\bibnamefont
  {Cerjan}}, \bibinfo {author} {\bibfnamefont {M.}~\bibnamefont {J\"urgensen}},
  \bibinfo {author} {\bibfnamefont {W.~A.}\ \bibnamefont {Benalcazar}},
  \bibinfo {author} {\bibfnamefont {S.}~\bibnamefont {Mukherjee}}, \ and\
  \bibinfo {author} {\bibfnamefont {M.~C.}\ \bibnamefont {Rechtsman}},\ }\textit{Observation of a Higher-Order Topological Bound State in the Continuum},  \href{\doibase 10.1103/PhysRevLett.125.213901} {\bibfield  {journal} {\bibinfo
  {journal} {Phys. Rev. Lett.}\ }\textbf {\bibinfo {volume} {125}},\ \bibinfo
  {pages} {213901} (\bibinfo {year} {2020})}\BibitemShut {NoStop}%
\bibitem [{\citenamefont {El~Hassan}\ \emph {et~al.}(2019)\citenamefont
  {El~Hassan}, \citenamefont {Kunst}, \citenamefont {Moritz}, \citenamefont
  {Andler}, \citenamefont {Bergholtz},\ and\ \citenamefont
  {Bourennane}}]{Hassan19NPho}%
  \BibitemOpen
  \bibfield  {author} {\bibinfo {author} {\bibfnamefont {A.}~\bibnamefont
  {El~Hassan}}, \bibinfo {author} {\bibfnamefont {F.~K.}\ \bibnamefont
  {Kunst}}, \bibinfo {author} {\bibfnamefont {A.}~\bibnamefont {Moritz}},
  \bibinfo {author} {\bibfnamefont {G.}~\bibnamefont {Andler}}, \bibinfo
  {author} {\bibfnamefont {E.~J.}\ \bibnamefont {Bergholtz}}, \ and\ \bibinfo
  {author} {\bibfnamefont {M.}~\bibnamefont {Bourennane}},\ }\textit{Corner states of light in photonic waveguides}, \href {\doibase
  10.1038/s41566-019-0519-y} {\bibfield  {journal} {\bibinfo  {journal} {Nat.
  Photonics}\ }\textbf {\bibinfo {volume} {13}},\ \bibinfo {pages} {697}
  (\bibinfo {year} {2019})}\BibitemShut {NoStop}%
\bibitem [{\citenamefont {Xue}\ \emph {et~al.}(2021)\citenamefont {Xue},
  \citenamefont {Wang}, \citenamefont {Huang}, \citenamefont {Cheng},
  \citenamefont {Yu}, \citenamefont {Foo}, \citenamefont {Zhao}, \citenamefont
  {Yang},\ and\ \citenamefont {Zhang}}]{Xue21arXiv}%
  \BibitemOpen
  \bibfield  {author} {\bibinfo {author} {\bibfnamefont {H.}~\bibnamefont
  {Xue}}, \bibinfo {author} {\bibfnamefont {Z.}~\bibnamefont {Wang}}, \bibinfo
  {author} {\bibfnamefont {Y.-X.}\ \bibnamefont {Huang}}, \bibinfo {author}
  {\bibfnamefont {Z.}~\bibnamefont {Cheng}}, \bibinfo {author} {\bibfnamefont
  {L.}~\bibnamefont {Yu}}, \bibinfo {author} {\bibfnamefont {Y.~X.}\
  \bibnamefont {Foo}}, \bibinfo {author} {\bibfnamefont {Y.~X.}\ \bibnamefont
  {Zhao}}, \bibinfo {author} {\bibfnamefont {S.~A.}\ \bibnamefont {Yang}}, \
  and\ \bibinfo {author} {\bibfnamefont {B.}~\bibnamefont {Zhang}},\
  }\textit{Projectively Enriched Symmetry and Topology in Acoustic Crystals}, \href{\doibase 10.1103/PhysRevLett.128.116802} {\bibfield  {journal} {\bibinfo
  {journal} {Phys. Rev. Lett.}\ }\textbf {\bibinfo {volume} {128}},\ \bibinfo
  {pages} {116802} (\bibinfo {year} {2022})}\BibitemShut {NoStop}%
\bibitem [{\citenamefont {Li}\ \emph {et~al.}(2021{\natexlab{b}})\citenamefont
  {Li}, \citenamefont {Du}, \citenamefont {Zhang}, \citenamefont {Li},
  \citenamefont {Fan}, \citenamefont {Zhang},\ and\ \citenamefont
  {Qiu}}]{QiuC21arxiv}%
  \BibitemOpen
  \bibfield  {author} {\bibinfo {author} {\bibfnamefont {T.}~\bibnamefont
  {Li}}, \bibinfo {author} {\bibfnamefont {J.}~\bibnamefont {Du}}, \bibinfo
  {author} {\bibfnamefont {Q.}~\bibnamefont {Zhang}}, \bibinfo {author}
  {\bibfnamefont {Y.}~\bibnamefont {Li}}, \bibinfo {author} {\bibfnamefont
  {X.}~\bibnamefont {Fan}}, \bibinfo {author} {\bibfnamefont {F.}~\bibnamefont
  {Zhang}}, \ and\ \bibinfo {author} {\bibfnamefont {C.}~\bibnamefont {Qiu}},\
  }\textit{Acoustic Möbius Insulators from Projective Symmetry}, \href{\doibase 10.1103/PhysRevLett.128.116803} {\bibfield  {journal} {\bibinfo
  {journal} {Phys. Rev. Lett.}\ }\textbf {\bibinfo {volume} {128}},\ \bibinfo
  {pages} {116803} (\bibinfo {year} {2022})}\BibitemShut {NoStop}%
\bibitem [{\citenamefont {Shao}\ \emph {et~al.}(2021)\citenamefont {Shao},
  \citenamefont {Liu}, \citenamefont {Xiao}, \citenamefont {Yang},\ and\
  \citenamefont {Zhao}}]{Shao21prl}%
  \BibitemOpen
  \bibfield  {author} {\bibinfo {author} {\bibfnamefont {L.~B.}\ \bibnamefont
  {Shao}}, \bibinfo {author} {\bibfnamefont {Q.}~\bibnamefont {Liu}}, \bibinfo
  {author} {\bibfnamefont {R.}~\bibnamefont {Xiao}}, \bibinfo {author}
  {\bibfnamefont {S.~A.}\ \bibnamefont {Yang}}, \ and\ \bibinfo {author}
  {\bibfnamefont {Y.~X.}\ \bibnamefont {Zhao}},\ }\textit{Gauge-Field Extended 
k.p Method and Novel Topological Phases},  \href {\doibase
  10.1103/PhysRevLett.127.076401} {\bibfield  {journal} {\bibinfo  {journal}
  {Phys. Rev. Lett.}\ }\textbf {\bibinfo {volume} {127}},\ \bibinfo {pages}
  {076401} (\bibinfo {year} {2021})}\BibitemShut {NoStop}%
\bibitem [{\citenamefont {Wakabayashi}\ \emph {et~al.}(1999)\citenamefont {Wakabayashi},
  \citenamefont {Fujita},\citenamefont {Ajiki},\ and\ \citenamefont {Sigrist}}]   {Wakabayashi99prb}%
  \BibitemOpen
  \bibfield  {author} {\bibinfo {author} {\bibfnamefont {K.}\ \bibnamefont
  {Wakabayashi}}, \bibinfo {author} {\bibfnamefont {M.}\ \bibnamefont
  {Fujita}}, \bibinfo {author} {\bibfnamefont {H.}\ \bibnamefont
  {Ajiki}}, \ and\ \bibinfo {author} {\bibfnamefont {M.}\ \bibnamefont
  {Sigrist}},\ }\textit{Electronic and magnetic properties of nanographite ribbons},  \href {\doibase 10.1103/PhysRevB.59.8271} {\bibfield  {journal}
  {\bibinfo  {journal} {Phys. Rev. B}\ }\textbf {\bibinfo {volume} {59}},\
  \bibinfo {pages} {8271} (\bibinfo {year} {1999})}\BibitemShut {NoStop}%
\bibitem [{\citenamefont {Mielke}(1992)}]{Mielke92JPA}%
  \BibitemOpen
  \bibfield  {author} {\bibinfo {author} {\bibfnamefont {A.}~\bibnamefont
  {Mielke}},\ }\textit{Exact ground states for the Hubbard model on the Kagome lattice},  \href {\doibase 10.1088/0305-4470/25/16/011} {\bibfield
  {journal} {\bibinfo  {journal} {Journal of Physics A: Mathematical and
  General}\ }\textbf {\bibinfo {volume} {25}},\ \bibinfo {pages} {4335}
  (\bibinfo {year} {1992})}\BibitemShut {NoStop}%
\bibitem [{\citenamefont {Yang}\ \emph {et~al.}(2022)\citenamefont {Yang},
  \citenamefont {Song}, \citenamefont {Cao},\ and\ \citenamefont {Yan}}]{Yang22nano}%
  \BibitemOpen
  \bibfield  {author} {\bibinfo {author} {\bibfnamefont {H.}\ \bibnamefont
  {Yang}}, \bibinfo {author} {\bibfnamefont {L.}\ \bibnamefont
  {Song}}, \bibinfo {author} {\bibfnamefont {Y.}\ \bibnamefont
  {Cao}},\ and\ \bibinfo {author} {\bibfnamefont {P.}\ \bibnamefont
  {Yan}},\ }\textit{Experimental Realization of Two-Dimensional Weak Topological Insulators},   \href {\doibase 10.1021/acs.nanolett.2c00555} {\bibfield  {journal}
  {\bibinfo  {journal} {Nano Lett.}\ }\textbf {\bibinfo {volume} {22(7)}},\
  \bibinfo {pages} {3125} (\bibinfo {year} {2022})}\BibitemShut {NoStop}%
\bibitem [{\citenamefont {Tam}\ \emph {et~al.}(2022)\citenamefont {Tam},
  \citenamefont {Venderbos},\ and\ \citenamefont {Kane}}]{Tam22prb}%
  \BibitemOpen
  \bibfield  {author} {\bibinfo {author} {\bibfnamefont {P.~M.}\ \bibnamefont
  {Tam}}, \bibinfo {author} {\bibfnamefont {J.~W.~F.}\ \bibnamefont
  {Venderbos}}, \ and\ \bibinfo {author} {\bibfnamefont {C.~L.}\ \bibnamefont
  {Kane}},\ }\textit{Toric-code insulator enriched by translation symmetry},   \href {\doibase 10.1103/PhysRevB.105.045106} {\bibfield  {journal}
  {\bibinfo  {journal} {Phys. Rev. B}\ }\textbf {\bibinfo {volume} {105}},\
  \bibinfo {pages} {045106} (\bibinfo {year} {2022})}\BibitemShut {NoStop}%
\end{thebibliography}
\end{document}